\documentclass[a4paper,fleqn,usenatbib]{mnras}

\usepackage{txfonts}

\usepackage[T1]{fontenc}
\usepackage{ae,aecompl}
\usepackage{graphicx}	% Including figure files
\usepackage{amssymb}	% Extra maths symbols
\usepackage{pdfpages}

\title[A new look inside LoTr 5]{A new look inside Planetary Nebula  LoTr 5: A long-period binary with hints of a possible third component }

\author[A. Aller et al.]{
A. Aller$^{1}$\thanks{E-mail: alba.aller@ifa.uv.cl},
J. Lillo-Box$^{2}$,
M. Vu\v{c}kovi\'{c}$^{1}$,
H. Van Winckel$^{3}$,
D. Jones$^{4,5}$,
\newauthor 
B. Montesinos$^{6}$,
M. Zorotovic$^{1}$,
and L. F. Miranda$^{7}$
\\
$^{1}$Instituto de F\'isica y Astronom\'ia, Facultad de Ciencias, Universidad de Valpara\'iso, Gran Breta\~na 1111, Playa Ancha, \\Valpara\'iso, 2360102, , Chile\\
$^{2}$European Southern Observatory (ESO), Alonso de Cordova 3107, Vitacura, Casilla 19001, Santiago de Chile, Chile\\
$^{3}$Instituut voor Sterrenkunde, KU Leuven, Celestijnenlaan 200D bus 2401, 3001, Leuven, Belgium\\
$^{4}$Instituto de Astrof\'isica de Canarias, 38205 La Laguna, Tenerife, Spain\\
$^{5}$Departamento de Astrof\'isica, Universidad de La Laguna, 38206 La Laguna, Tenerife, Spain\\
$^{6}$Departamento de Astrof\'{\i}sica, Centro de Astrobiolog\'{\i}a (INTA-CSIC), PO Box 78, E-28691 Villanueva de la Ca\~nada (Madrid), Spain\\
$^{7}$Instituto de Astrof\'{\i}sica de Andaluc\'{\i}a - CSIC, C/ Glorieta de la Astronom\'{i}a s/n, E-18008 Granada, Spain
}

\date{Accepted XXX. Received YYY; in original form ZZZ}

\pubyear{2015}

\begin{document}
\label{firstpage}
\pagerange{\pageref{firstpage}--\pageref{lastpage}}
\maketitle

% Abstract of the paper
\begin{abstract}
LoTr 5 is a planetary nebula with an unusual long-period binary central star. As far as we know, the pair consists of a rapidly rotating G-type star and a hot star, which is responsible for the ionization of the nebula. The rotation period of the G-type star is 5.95 days and the orbital period of the binary is now known to be $\sim$2700 days, one of the longest in central star of planetary nebulae. The spectrum of the G central star shows a complex H$\alpha$ double-peaked profile which varies with very short time scales, also reported in other central stars of planetary nebulae and whose origin is still unknown. We present new radial velocity observations of the central star which allow us to confirm the orbital period for the long-period binary and discuss the possibility of a third component in the system at $\sim$129 days to the G star. This is complemented with the analysis of archival light curves from SuperWASP, ASAS and OMC. From the
spectral fitting of the G-type star, we obtain a effective temperature of $T_{\rm eff}$ = 5410$\pm$250\,K and surface gravity of $\log g$
= 2.7$\pm$0.5, consistent with both giant and subgiant stars. We also present a detailed analysis of the H$\alpha$ double-peaked profile and conclude that it does not present correlation with the rotation period and that the presence of an accretion disk via Roche lobe overflow is unlikely.\end{abstract}

% Select between one and six entries from the list of approved keywords.
% Don't make up new ones.

%%%%%%%%%%%%%%%%%%%%%%%%%%%%%%%%%%%%%%%%%%%%%%%%%%

%%%%%%%%%%%%%%%%% BODY OF PAPER %%%%%%%%%%%%%%%%%%

\section{Saga of LoTr 5}
\label{sec:introduction} 

\begin{figure}
	\includegraphics[width=\columnwidth]{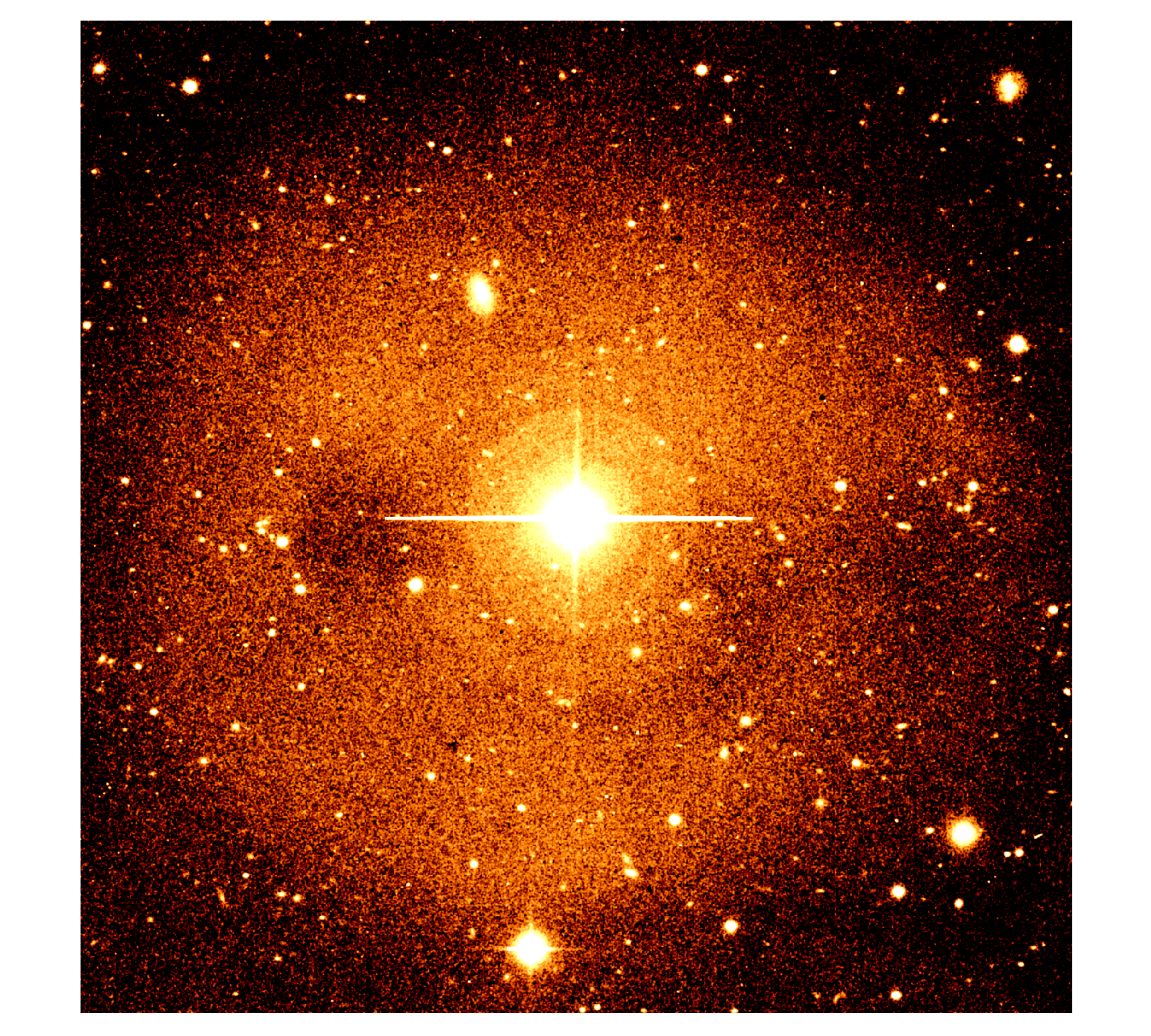}
    \caption{Image of LoTr 5 taken with the Isaac Newton Telescope's Wide
Field Camera in the H$\alpha$ filter. The size of the field is 9\arcmin  $\times$ 9\arcmin.  North is up and
East is left.}
    \label{fig:image_LoTr5}
\end{figure}

 The planetary nebula (PN) LoTr 5 ($\alpha_{(2000.0)}$ = 12$^h$\,55$^m$\,33$\fs$7, $\delta_{(2000.0)}$
= $+$25$^{\circ}$\,53$'$\,30$''$; see Figure\,\ref{fig:image_LoTr5}) was discovered by \cite{Longmore-Tritton1980}. It is the highest Galactic latitude PN in our Galaxy (b= +88$^{\circ}$.5) and belongs to the so-called Abell 35-type\footnote{As already discussed in \cite{Tyndall2013}, we consider that the Abell 35-type PNe designation is not longer appropriate, since Abell 35 is most likely a Str\"omgren zone in the ambient interstellar medium \citep{Frew2008} and not a true PN.} group of PNe \citep{Bond1993}. The PNe in this group are known to host a binary central star comprising a rapidly rotating giant or subgiant star plus a hot progenitor responsible for the ionization of the nebula. These objects have some resemblance to other long-period binaries as, e.g., the high-galactic-latitude HD 128220 \citep[see][and references therein]{Heber2016} in which possible associated PNe would have already dissipated. The optical spectrum of HD\,112313, the central star of LoTr 5 (also named IN Comae), is completely dominated by a rapidly rotating G-type star. The evolutionary status of this star is uncertain and both giant and subgiant solutions have been suggested in the literature \citep{Strassmeier1997}. The hot component was firstly found by \cite{Feibelman-Kaler1983} by analyzing ultraviolet spectra, in which the flux peak showed an extremely hot (over 100\,000\,K) component. Since then, searches for long- and short-period variability have been persistently done, culminating in the recent detection of a long orbital period binary \citep[][]{VanWinckel2014,Jones2017}.

Photometric variability in HD\,112313 was firstly found by \cite{Schnell-Purgathofer1983}, who reported four different periods between 0.35 and 1.2 days, although they stated that the 1.2-day period was the most probable. Later on, \cite{Noskova1989} detected the 5.9 days period, which was independently found by H. E. Bond in 1988 \citep[see][]{Bond-Livio1990}. This 5.9 days period was also confirmed in the subsequent years by \cite{Kuczawska-Mikolajewski1993}, \cite{Jasniewicz1996}, and \cite{Strassmeier1997}, who interpret this period as the rotational period of the G-type star and the 1.2 days period as an alias of the true period. This is still assumed today. 

Parallel to these works, radial velocity variations were also further discussed. \cite{Acker1985} were the first to detect radial velocity variations in HD\,112313 and they derived a probable (but questionable due to imprecision of the observations) period of 0.35 days. Two years later, \cite{Jasniewicz1987}, with {\sc coravel} high-resolution spectra, proposed a triple scenario for the system, consisting of an inner close-binary of 1.99 days orbital period (with two similar components), and an outer, third companion (the hot sdO star) with $\sim$540 days orbital period. \cite{Malasan1991} also concluded the triple scenario, but with a different configuration: a 1.75 days period for the inner system and 2000 days for the outer one, being doubtful whether the sdO belongs to the inner or outer orbit. \cite{Jasniewicz1994}, apart from photometric variations, also detected radial velocity variations although no stable period was found due to the poor quality of the spectra. \cite{Strassmeier1997} also obtained radial velocity variations but they concluded that they were due to the influence of starspots. The periodicity of several years was confirmed by \cite{VanWinckel2014}, although they did not cover the full radial velocity curve. After all of these efforts, the orbital period has been recently unveiled by \cite{Jones2017}, being one of the longest period measured in a central star of a PN \citep[$\sim$2700 days; see][for comparison with the period distribution of all the binary central stars]{Jones-Boffin2017}. \cite{Ciardullo1999} obtained high-resolution images of the nucleus of LoTr 5 with the Hubble Space Telescope but failed to resolve the binary, although this might be due to the very small angular separation
of the binary and/or to the faintness of the hot star.

Even though the orbit has already fully covered, there is still no clear scenario for this peculiar system. From high-resolution, long-slit spectra of the nebula, \cite{Graham2004} proposed a bipolar model for LoTr 5, in which the bipolar axis is tilted by $\sim$17$^{\circ}$ to the line of sight. However, that inclination is at odds with the mass function derived by \cite{Jones2017}, who proposed several explanations to solve the mass problem. One possibility is that the inclination of the nebula is incorrect. This solution seems to be the most probable since, a small change in the inclination would resolve the mass problem: for example, for inclinations > 23$^{\circ}$, the mass of the primary would be less than the Chandrasekhar limit (1.4 M$\odot$). Another possibility is that the bipolar axis of the nebula is not perpendicular to the orbital plane of the binary. 

 It is also worth to mention that LoTr 5 is one of the PNe with a 'Barium star' central star \citep{Miszalski2013b, Tyndall2013}. Barium stars are characterized by a strong enhancement of the barium abundance and other s-process elements such as Sr and Y \citep{Bidelman-Keenan1951}, as already showed by \cite{Thevenin-Jasniewicz1997} in their spectral analysis of the G-star of LoTr 5. This could be consequence of the contamination of the photosphere by the transfer material of s-process-overabundant material from the former AGB star (which is now the central star). This contamination process is thought to occur via wind-accretion and not through a common envelope evolution \citep{Boffin-Jorissen1988, Jeffries-Stevens1996}, which is in agreement with the long orbital period derived for the system \citep[see][]{Jones2017}.

Another remarkable aspect of HD\,112313 is its H$\alpha$ double-peaked profile that varies with very short time scales. This profile was firstly reported by \cite{Jasniewicz1994} and the origin is still unknown. The main causes proposed to explain it have been the presence of an accretion disc, the chromospheric activity of the G star, and/or stellar winds. Regarding the last point, \cite{Modigliani1993} reported a fast wind speed of 3300 km\,s$^{-1}$ from the IUE spectra, although the evidence for a such wind was questioned by \cite{Montez2010}. Finally, HD\,112313 was also confirmed as an X-rays emitter by \cite{Apparao1992} with the EXOSAT satellite. Later, \cite{Montez2010} detected point source X-ray emission of HD\,112313 with XMM and CXO spectra, stating that the most likely scenario is  the presence of coronal activity associated with the G-type star.

In this work, we revisit this complex system and present new radial velocity data of HD\,112313, which allow us to confirm the orbital period for the long-period binary and examine the possibility of a third component in the system close to the G-type star. These new data are complemented with the analysis of the archival light curves from SuperWasp, OMC and ASAS, that have not been published to date. In addition, a detailed analysis of the H$\alpha$ double-peaked profile is presented as well as the analysis of the IUE spectra available for this object.

\section{Observations and data reduction}

 %=======================================
\subsection{High-resolution, optical spectroscopy}
%=======================================
\label{sec:cafe} 

We used high-resolution spectra of  HD\,112313 from both archival and new observations obtained from different facilities.

We performed dedicated observations of this target using the Calar Alto Fiber-fed Echelle spectrograph \citep[CAFE,][]{Aceituno2013}  mounted on the 2.2-m telescope at Calar Alto Observatory (Almer\'{i}a, Spain).  CAFE covers the 4000--9500 $\AA$ spectral range with an average spectral resolution of $\lambda$/$\Delta$$\lambda$ = 63\,000. We acquired 6 high-resolution spectra on 2012 May 18-22.
The exposure times were fixed to 900\,s, with the signal-to-noise ratio ranging from 20 to 80 depending on the weather and seeing conditions, and one noisy spectra having S/N$<$10. The spectra were reduced with observatory's pipeline described in \cite{Aceituno2013}, which performs the standard routines for echelle spectroscopy: bias subtraction, order tracing, flat field correction, and wavelength calibration based on ThAr lamp frames obtained right after each science frame. 

\begin{figure}
	\includegraphics[width=\columnwidth]{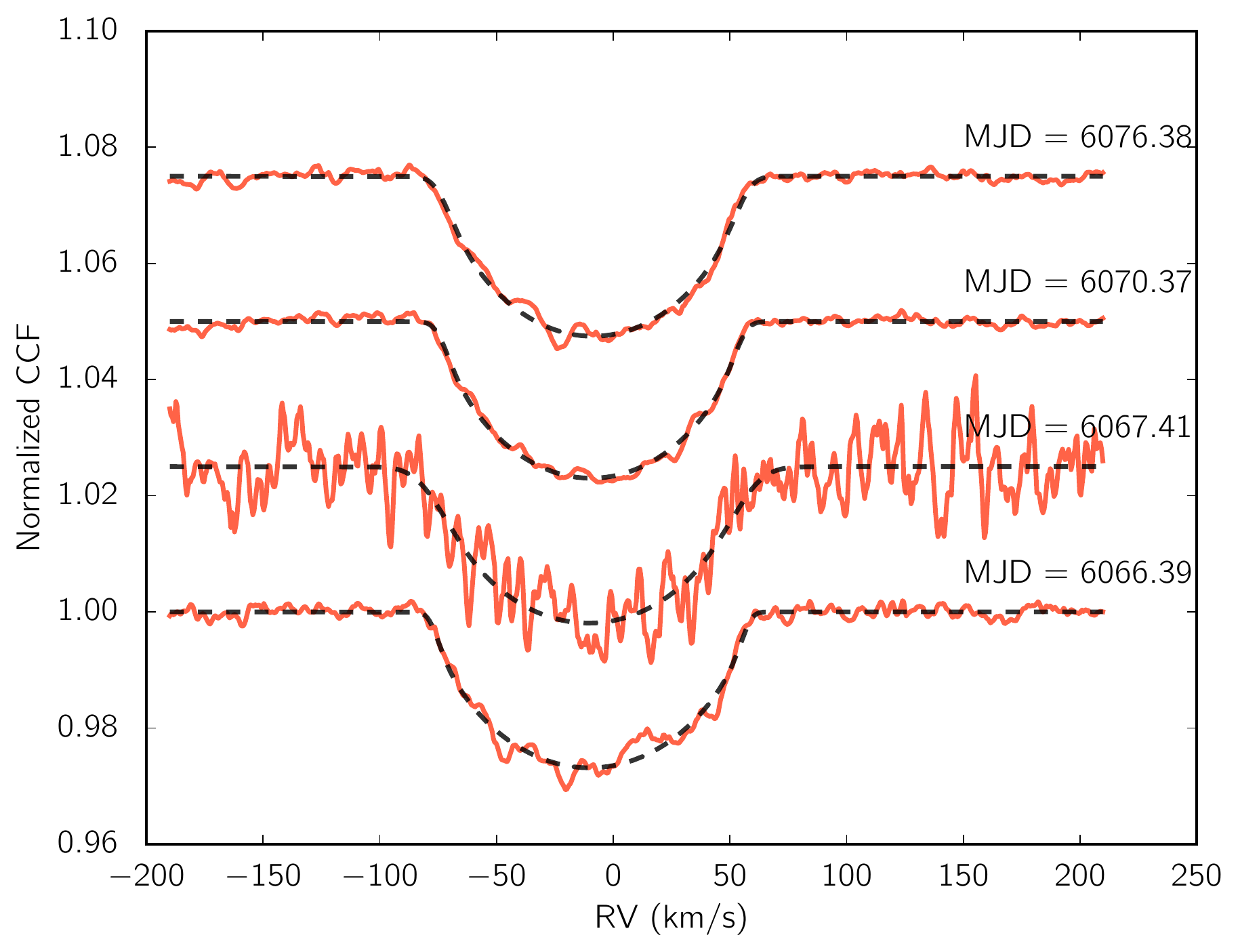}
    \caption{Cross-correlation functions of the CAFE spectra (solid red line) and their fitted profiles (dashed black line). The modified julian date (MJD) is indicated in each plot as JD-2450000. }
    \label{fig:CCFs_CAFE}
\end{figure}

The radial velocity was extracted from the individual spectra by cross-correlating them against a G2V binary mask \citep{Baranne1996} specially designed for the CAFE data and already used in planet-detection studies \citep[e.g., ][]{Lillo-Box2015}. Given the fast-rotating nature of the G-type star  ($v\sin{i}\sim 66.2\pm0.7$~km\,s$^{-1}$, \citealt{VanWinckel2014}), the cross-correlation function (CCF) was fitted to a rotational profile defined by \cite{Gray2005} and already used in planet-related studies like \cite{Santerne2012}

\begin{equation}
G(v) = \frac{2(1-\epsilon)\sqrt{1-(v/v_L)^2}+0.5\pi\epsilon \left[1-(v/v_L)^2\right]}{\pi v_L (1-\epsilon/3)}
\end{equation}

\noindent where $v_L=v\sin{i}$ and $\epsilon$ is a limb-darkening coefficient, that we fix to $\epsilon=0.3$ for the CCF modeling purposes in this paper as done in \cite{Santerne2012}. We note that this coefficient is not critical since radial velocities just change within the uncertainties for different values of $\epsilon$. This profile was convolved with the instrumental profile of CAFE, that we assume to be a Gaussian with a FWHM$=8.3$~km\,s$^{-1}$. The results of the CCF fitting are presented in Figure~\ref{fig:CCFs_CAFE} and the radial velocities with their corresponding uncertainties are summarized in Table~\ref{Table:RVs}. The CAFE radial velocity data taken in the same night have been combined to increase the precision. 

In addition, we retrieved the high-resolution echelle spectra of HD\,112313 from the archive of the ELODIE\footnote{\url{http://atlas.obs-hp.fr/elodie/}}  \citep{Moultaka2004} spectrograph at the Observatoire de Haute-Provence 1.93-m telescope. The spectra cover a spectral range of 4000--6800 $\AA$ with a resolution of R = 42000. For HD 112313, the ELODIE archive provides three epochs: two with exposure times of 3600\,s and S/N of 128 and 76, respectively, taken in 1998 March 7-8; and another with 900\,s and a S/N of 23, taken in 2003 March 11. The ELODIE archive also provides the radial velocity of each spectrum calculated by cross-correlating it with numerical masks of F0 or K0 spectral types. However, these radial velocities are obtained by fitting the CCF with a Gaussian profile, which is not suitable in this case due to the broad CCF given the fast rotation of the star. For that reason, we have re-calculated the radial velocity values by using the G2V binary mask and fitting the CCF with the same rotational profile used in the CAFE data. The radial velocities with their corresponding uncertainties are shown in Table~\ref{Table:RVs}.

\begin{table}
 \centering
  \caption{Radial velocity measurements from CAFE and ELODIE used in the analysis.}
  \begin{tabular}{@{}ccc@{}}
  \hline    
 Julian date 		& RV (km s$^{-1}$)		& Instrument  \\
  \hline \hline
2450879.090356 &-14.80$\pm$0.28 &ELODIE \\
2450880.0942862 &-15.64$\pm$0.33 &ELODIE \\
2452709.9643882 &-9.20$\pm$0.45 &ELODIE  \\
2456066.39055	&    -10.062$\pm$0.17 	& CAFE\\
2456067.40529	&	 -10.52$\pm$0.75 & CAFE    \\
2456070.37146	& -9.45$\pm$0.11  	& CAFE   \\
2456076.37568 	& -8.75$\pm$0.13   	& CAFE \\

\hline
\end{tabular}
\label{Table:RVs}
\end{table}

Finally, we have also used the data published in \cite{Jones2017} for this object, taken with the HERMES spectrograph \citep{Raskin2011} on the 1.2m Mercator telescope, as part of a large radial-velocity monitoring programme. The data consist of 210 radial velocity measurements with a very high precision that have been critical to confirm the long-period binary star. The radial velocities from HERMES can be found in \cite{Jones2017}.

\begin{figure}
\includegraphics[width=\columnwidth]{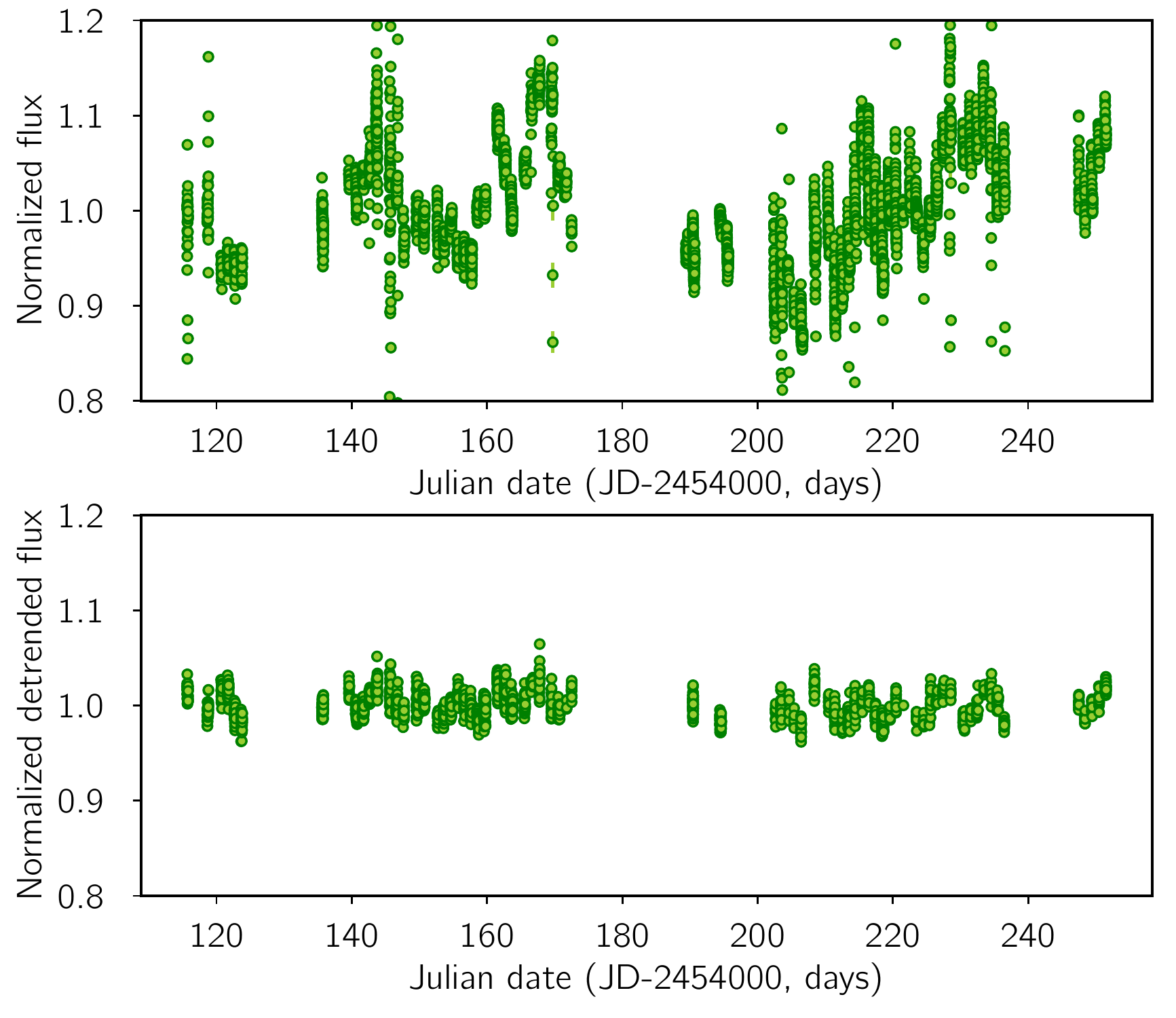}
\caption{SuperWASP light curve. Top: undetrended flux from the superWASP archive. Bottom: Detrended flux by using the close-by star HD 112299 as reference.}
\label{fig:detrending}
\end{figure}

\subsection{Archival photometry}

We retrieved the photometric time series for HD\,112313 from the surveys SuperWASP \citep[Wide Angle Search for
Planets;][]{Pollacco2006}, ASAS-3 (All Sky Automated Survey) Photometric V-band Catalogue \citep{Pojmanski1997} and OMC \citep[Integral's Optical Monitoring Camera;][]{Mas-Hesse2003}. 

The SuperWASP monitoring covers 116 days with 5323 data points taken with a broad band passband from 400 to 700 nm. We retrieved the data from the SuperWASP Public Archive\footnote{\url{http://wasp.cerit-sc.cz/form}}. Given the large instrumental imprints in the flux, we also retrieved the photometric data from a star close to the target in order to remove these drifts in a precise manner. We then performed differential photometry by using HD 112299 as a reference star. The original and detrended light curves are shown in Fig.~\ref{fig:detrending}. The OMC data, taken in the Johnson V-band, were retrieved from the OMC Archive\footnote{\url{https://sdc.cab.inta-csic.es/omc/secure/form_busqueda.jsp}} and provide a time span of the observations of 3646 days with 2745 data points. Here we have cleaned the light curve following the criteria in \cite{Alfonso-Garzon2012} with the aim of improving the quality of the data. In the case of the ASAS data, retrieved from the ASAS All Star Catalogue\footnote{\url{http://www.astrouw.edu.pl/asas/?page=aasc&catsrc=asas3}}, we only have 272 data points in 2370 days time span. ASAS provides photometry with 5 different apertures. For stars with magnitudes brighter than 9 (as it is the case of HD\,112313), it is recommended to use the one with the widest aperture ($MAG4$\footnote{\url{http://www.astrouw.edu.pl/asas/explanations.html}}).

The Optical Monitoring Camera (OMC) on board the high-energy INTEGRAL satellite provides photometry in the Johnson V-band within a 5 by 5 degree field of view. 

\subsection{Low resolution ultraviolet spectra}

 We have retrieved all the IUE \citep[International Ultraviolet Explorer;][]{Kondo1989} spectra available in the IUE Newly Extracted Spectra
(INES\footnote{\url{http://sdc.cab.inta-csic.es/ines/}}) System for HD\,112313. INES provides spectra already calibrated in physical units.

We have retrieved the available low-dispersion ($\sim$ 6\,$\AA$) IUE
SWP (short wavelength) and LWP (long wavelength) spectra, which were
obtained from 1982 to 1990. The SWP and LWP spectra cover the ranges 
1150-1975\,$\AA$ and 1910-3300\,$\AA$, respectively. In total, 28 SWP and LWP spectra are available. They are listed in Table~\ref{Table:IUEspectra}, with the exposure times and the date of the observations.  We discarded the spectra with many bad pixels and/or those that appear to be saturated.

\begin{table}
 \centering
  \caption{Summary of the IUE observations available in the INES database for HD\,112313.}
  \begin{tabular}{@{}lcc@{}}
  \hline    
 Spectrum & Exp (s) & Date \\%& Comments\\
  \hline \hline
 SWP16896 	&  	1200			&	1982-05-05   	\\%&	Bad pixels below 1450$\AA$\\
 SWP16897 	&  	1200			&	1982-05-05   	\\%&	Bad pixels above 2400$\AA$   \\
 LWR13173	&  	1200			&	1982-05-05   	\\%& 	Bad pixels below 1350$\AA$ \\
 SWP17236	&  	600			&	1982-06-16   	\\%&	Some isolated bad pixels all over the spectrum   \\
 LWR13502	&  	600			&	1982-06-16   	\\%&	Some isolated bad pixels all over the spectrum  \\
 LWR15882	&  	2100			&	1983-05-05   	\\%&  	High flux errors \\
SWP19909	& 	300			&	1983-05-05   	\\%&	Some isolated bad pixels all over the spectrum  \\
 LWR15883 	& 	300			&	1983-05-05   	\\%&	Some isolated bad pixels all over the spectrum and high flux errors in some areas\\
 LWR15884	& 	5100			&	1983-05-05   	\\%&	High flux errors \\
 SWP35635 	& 	480			&	1989-02-28   	\\%&	Some isolated bad pixels all over the spectrum \\
 SWP35643 	& 	900			&	1989-03-01   	\\%& 	Some isolated bad pixels all over the spectrum \\
 SWP35688	& 	1200			&	1989-03-06   	\\%&	Bad pixels below 1350$\AA$ \\
 LWP15137 	& 	1200			&	1989-03-06   	\\%&	Some isolated bad pixels all over the spectrum (specially in the MgII line) \\
 SWP35709	& 	1200			&	1989-03-08   	\\%&	Bad pixels below 1400$\AA$ \\
 LWP17022 	& 	1080			&	1989-12-28   	\\%&	Some isolated bad pixels all over the spectrum (specially in the MgII line) \\
 SWP37912 	&  	900			&	1989-12-28   	\\%&	Bad pixels below 1350$\AA$ \\
 SWP37912 	&  	1200			&	1989-12-28   	\\%&	Some isolated bad pixels all over the spectrum   \\
 LWP17023	&  	480			&	1989-12-28   	\\%&	Some isolated bad pixels all over the spectrum  \\
 LWP17042	&  	720			&	1989-12-30   	\\%&	Bad pixels below 2250$\AA$   \\
SWP37923	& 	900			&	1989-12-30   	\\%&	Bad pixels below 1350$\AA$ \\
 LWP17063 	& 	600			&	1990-01-01   	\\%&	Some isolated bad pixels all over the spectrum   \\
 LWP17063	& 	600			&	1990-01-01   	\\%&	Some isolated bad pixels all over the spectrum   \\
 SWP37932 	& 	900			&	1990-01-01   	\\%&	Bad pixels below 1350$\AA$  \\
 SWP37936 	& 	900			&	1990-01-02   	\\%&	Bad pixels below 1350$\AA$  \\
 LWP17070 	& 	720			&	1990-01-02   	\\%&	Some isolated bad pixels all over the spectrum   \\
 LWP17070 	& 	600			&	1990-01-02   	\\%&	Some isolated bad pixels all over the spectrum  and high flux errors below 2250$\AA$ \\
 SWP37937	& 	20280		&	1990-01-02   	\\%&	Some isolated bad pixels all over the spectrum    \\
 SWP38776 	& 	20400		&	1990-05-12   	\\%&	Some isolated bad pixels all over the spectrum \\
\hline
\end{tabular}
\label{Table:IUEspectra}
\end{table}

 \begin{table*}
 \centering
  \caption{Results from the light curve analysis of the three datasets.}
    \begin{tabular}{lccc}

  \hline    
Parameter	& 	SuperWASP  &  ASAS & OMC\\
  \hline \hline
Period 	(days)	& 	5.9508$\pm$0.0006  &  5.9670$\pm$0.0014	& 5.9702$\pm$0.0002 \\
$\phi_0$ 	(days)		&  	2452652.01$\pm$0.16& 2452652.30$\pm$0.21	 & 2452651.951$\pm$0.043	\\
Amplitud 	(mmag)	&  	12.93$\pm$0.52	      & 20.2$\pm$2.9 		          & 18.12$\pm$0.41\\
Date interval 		&	2007 Jan -- 2007 May    &		2003 June -- July 2009	&		2003 Jan -- 2013 Jan		\\
Number  of data points &  5323					& 	242			&  2745 \\	
\hline
\end{tabular}
\label{Table:results_LCs}
\end{table*}

\section{Results}

\subsection{Spectral fitting of the G-type secondary}

The exact spectral type and evolutionary status of the G-type companion is uncertain, with both giant and subgiant solutions being possible \citep{Strassmeier1997}. In order to attempt to resolve this ambiguity, we fit the high signal-to-noise HERMES spectra using the spectral synthesis and modelling tool iSpec \citep{Blanco-Cuaresma2014}.  MARCS model atmospheres \citep{Gustafsson2008} were used to produce synthetic spectra using the \textsc{spectrum} spectral synthesis code \citep{Gray1994}, which were then iteratively fit to the observed spectrum following the chi-squared minimisation routine outlined in \citep{Blanco-Cuaresma2014}. The effective temperature, surface gravity and metallicity were allowed to vary, while the rotational broadening was fixed to a value of v sin i = 66.2 km\,s$^{-1}$ as determined by \citet{VanWinckel2014}.  The resulting effective temperature, $T_{\rm eff}$ = 5410$\pm$250\,K, and surface gravity, $\log g$ =2.7$\pm$0.5, are consistent with both giant and subgiant possible companions, while the low derived metallicity, [Fe/H] = -0.1$\pm$0.1 \citep[assuming solar abundances from][]{Grevesse2007}, is consistent with the high galactic latitude of LoTr 5.

\begin{figure}
\includegraphics[width=\columnwidth]{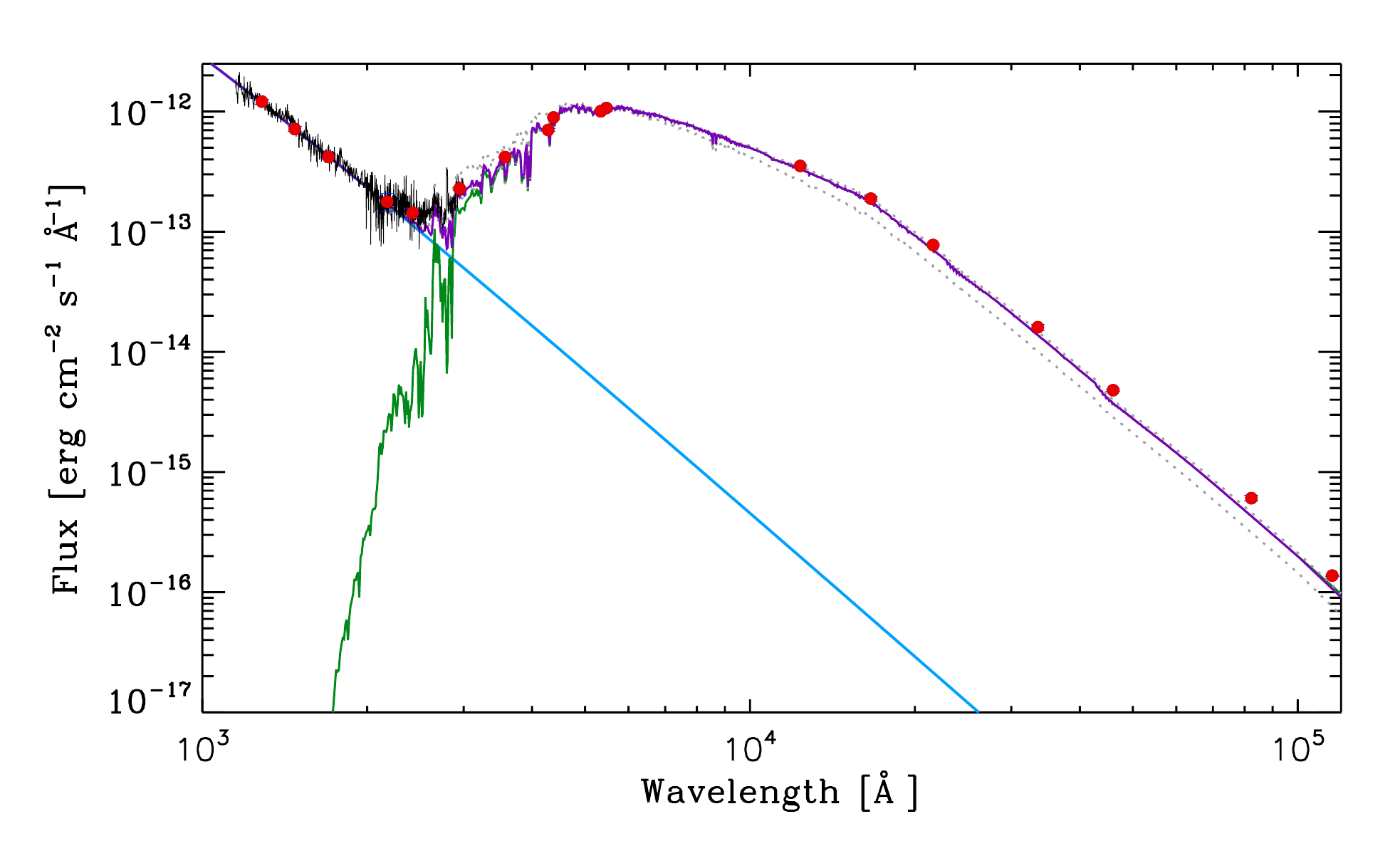}
\caption{Spectral energy distribution (SED) of HD\,112313 constructed with VOSA. One IUE spectrum is also included. The blue line corresponds to a blackbody with {\it T}$_{\rm eff}$ = 150000\,K. The green line corresponds to a Kurucz model with {\it T}$_{\rm eff}$ = 5410\,K and $\log g$
= 2.70. In purple is the composite model. The dotted grey lines represent the models with {\it T}$_{\rm eff}$ = 5160\,K and {\it T}$_{\rm eff}$ = 5660\,K, corresponding to the 1-sigma boundaries of the estimated effective temperature of LoTr 5.}
\label{fig:sed}
\end{figure}

\subsection{Spectral energy distribution}
\label{section:SED}
We have analyzed the spectral energy distribution (SED) of HD\,112313 by using the Virtual Observatory SED Analyzer \citep[VOSA,][]{Bayo2008}, in order to search for infrared excess. Figure\,\ref{fig:sed} presents the SED with the magnitudes from Tycho, UBV from Mermilliod 1991, 2MASS Point Source Catalog, and WISE archive. One IUE spectrum has been superimposed in the SED. In green, we have plotted a Kurucz model for the G-type star with {\it T}$_{\rm eff}$ = 5410\,K and $\log g$ = 2.70 and in blue a blackbody with {\it T}$_{\rm eff}$ = 150000\,K, to account for the hot star. The blackbody has not been reddened since the high Galactic latitude of LoTr 5 makes the extinction practically negligible. The planetary nebula is very faint so, apparently, there is not contamination from the nebula either. As can be seen in Fig.\,\ref{fig:sed}, no significant infrared excess can be identified, which would point out that there is no evidence for additional components (e.g., a third companion and/or an accretion disk) in the system. We will discuss in detail these possibilities below.

 %==============================
\subsection{Archival light curves analysis}
%==============================
\label{sec:lc}

\begin{figure}
\includegraphics[width=\columnwidth]{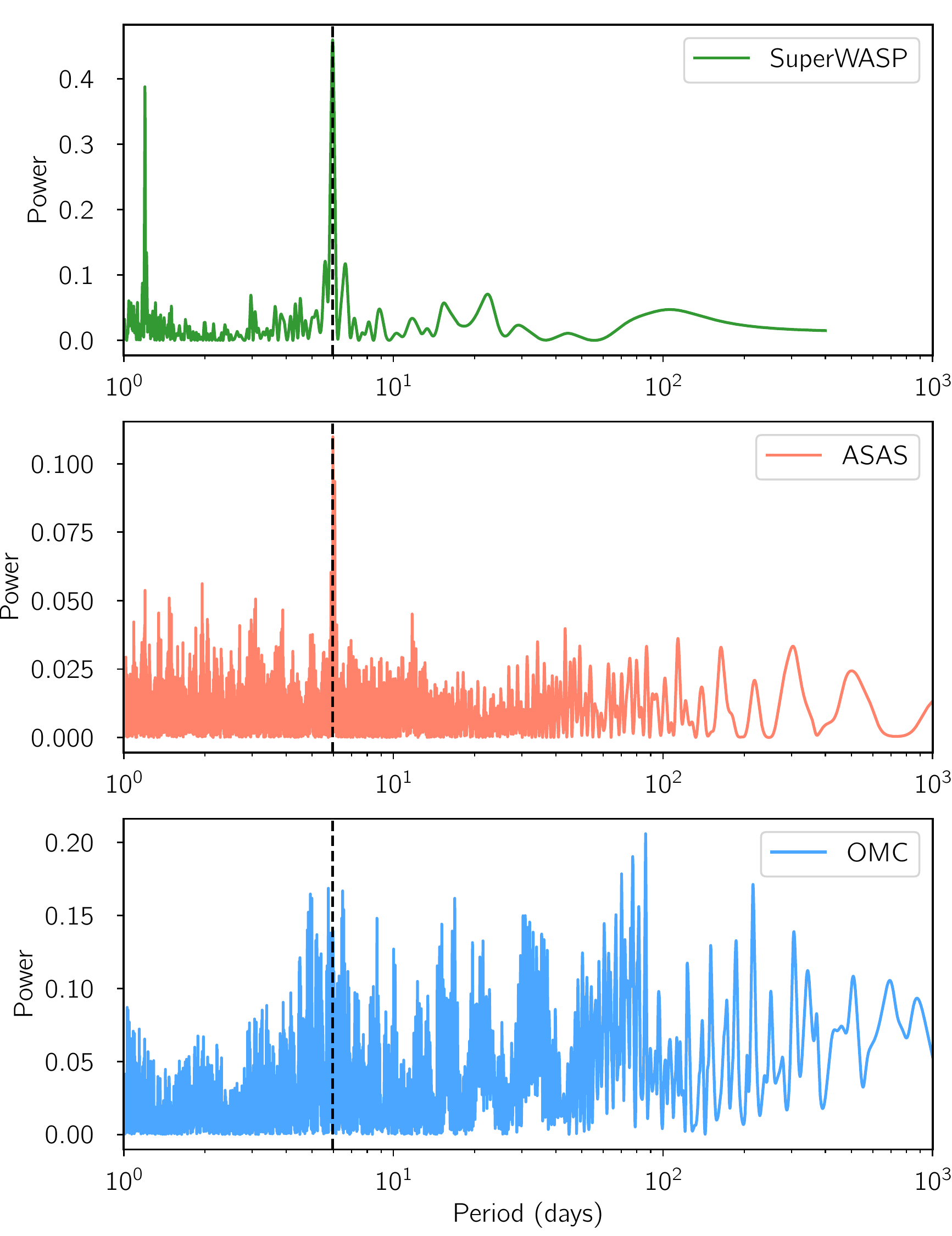}
\caption{Lomb-Scargle periodogram of the light curves provided by the three datasets used. The dashed vertical line indicates the $\sim5.95$~days period.}
\label{fig:LCperiodograms}
\end{figure}

\begin{figure}
\includegraphics[width=\columnwidth]{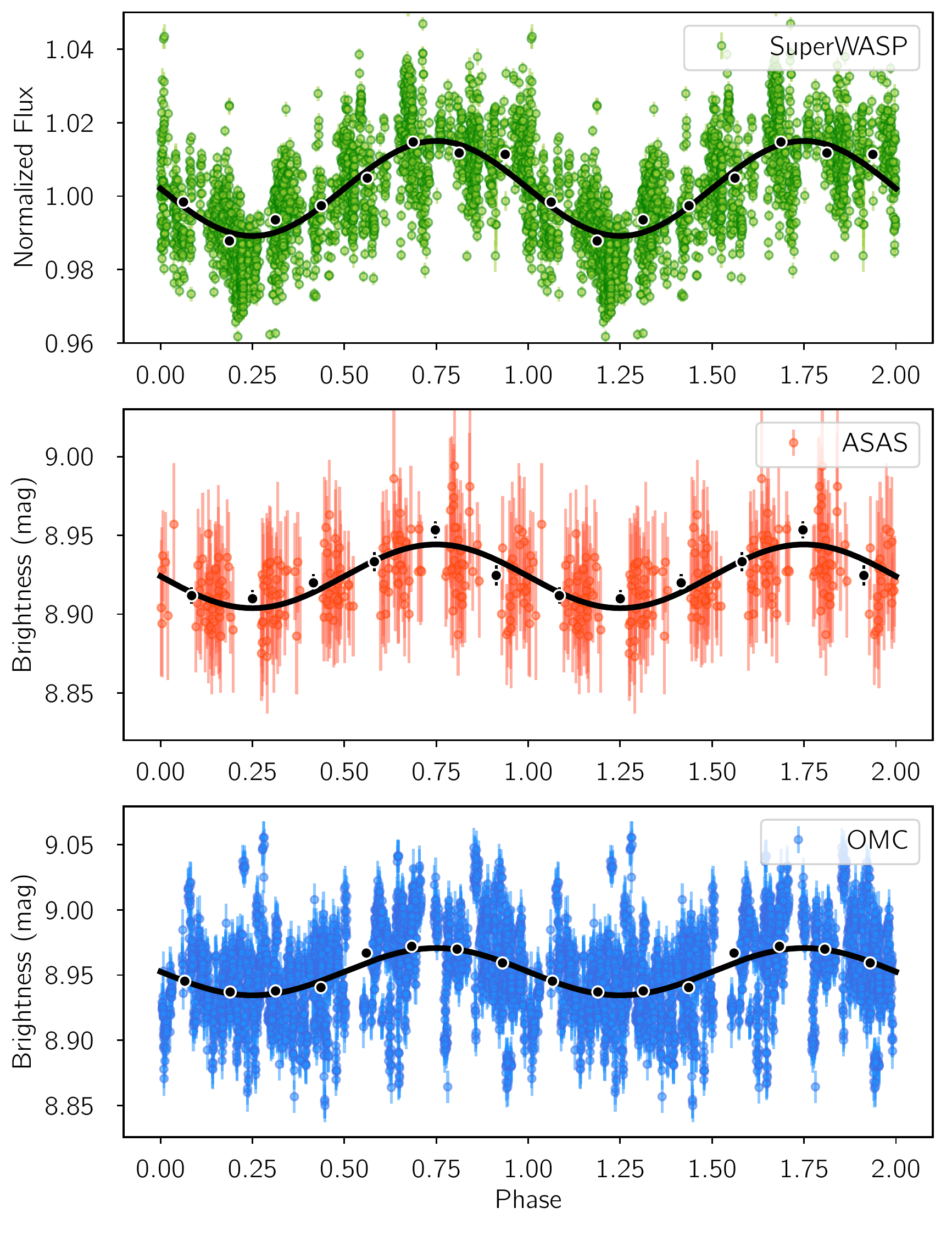}
\caption{Phase-folded light curves with the best fitted period for each of the three datasets, namely SuperWASP (top), ASAS (middle), and OMC (bottom). The black line shows the fitted sinusoidal model and the black circles represent the binned data.}
\label{fig:activity_wasp}
\end{figure}

In Fig.~\ref{fig:LCperiodograms}, we show the Lomb-Scargle periodograms of the three photometric datasets. In the case of superWASP and ASAS, we find a significant peak at $P\sim5.95$~days (assumed to be the rotation period of the (sub)giant star, see Sect.~\ref{sec:introduction}). We can model this signal with a simple sinusoid to estimate the period and amplitude of the variations independently in each dataset. We left the period to vary freely between 0 and 10 days in all cases and found: $P_{\rm WASP} = 5.95092\pm0.00062$~days, $P_{\rm ASAS} = 5.9409\pm0.0021$~days, $P_{\rm OMC} = 5.94579\pm0.00012$~days. Fig.~\ref{fig:activity_wasp} shows the phase-folded light curves of the three datasets with these periods. The corresponding semi-amplitudes are $A_{\rm WASP}=12.93\pm0.52$~mmag, $A_{\rm ASAS}=20.2\pm2.9$~mmag, $A_{\rm OMC}=18.12\pm0.41$~mmag. These results, together with the date range of the data, are summarized in Table~\ref{Table:results_LCs}. The amplitude clearly varies from one lightcurve to each others. This might be due to the different timespan of the datasets covering different time ranges of the stellar cycle (i.e., different amount of star spots). In the case of SuperWASP, the data were taken in a short time span (only four months), so the stellar spots are steady along this time. On contrary, in the case of ASAS and OMC, where the data are spread over years, the amplitude due to star sports is diluted because we are probably mixing periods of stellar maxima and stellar minima.

\begin{figure}
\includegraphics[width=\columnwidth]{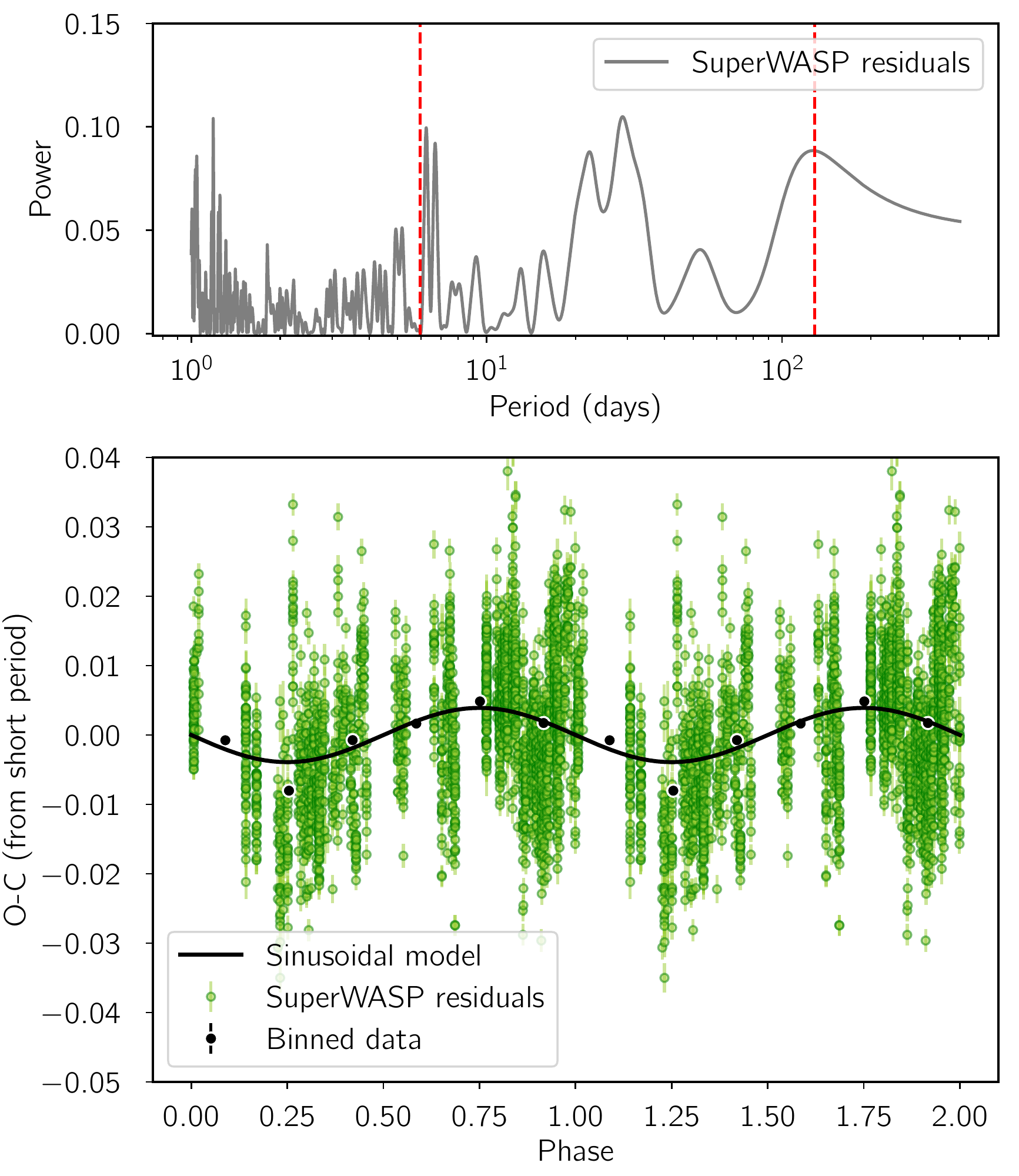}
\caption{Top: Periodogram of the residuals of the superWASP data after removing the $\sim$5.95~days signal. The $\sim129$~days peak is also marked with a vertical dashed line. Bottom: Light curve residuals after removing the$\sim$5.95~days signal and phase-folded with the 129 days periodicity}
\label{fig:superwasp_residuals}
\end{figure}

If we now remove this signal from the SuperWASP dataset (the most precise one) and calculate the periodogram of the residuals, other prominent (although non-significant) peaks appear (see Fig.~\ref{fig:superwasp_residuals}, top panel). In particular, we want to highlight the peak at $\sim129$~days. This is relevant because it also appears in the radial velocity analysis (see Sect.~\ref{sec:129days}). The phase-folded light curve with that period is shown in Fig.~\ref{fig:superwasp_residuals}, bottom panel. The origin of this peak will be discussed in detail in Sect.~\ref{sec:radial_velocity} together with the $\sim5.95$~days periodicity.

 %==============================
\subsection{Radial velocity analysis}
%==============================
\label{sec:radial_velocity}

%----------------------------------------------------
\subsubsection{The long-term companion}
%----------------------------------------------------
\label{sec:long}

\begin{figure}
\includegraphics[width=\columnwidth]{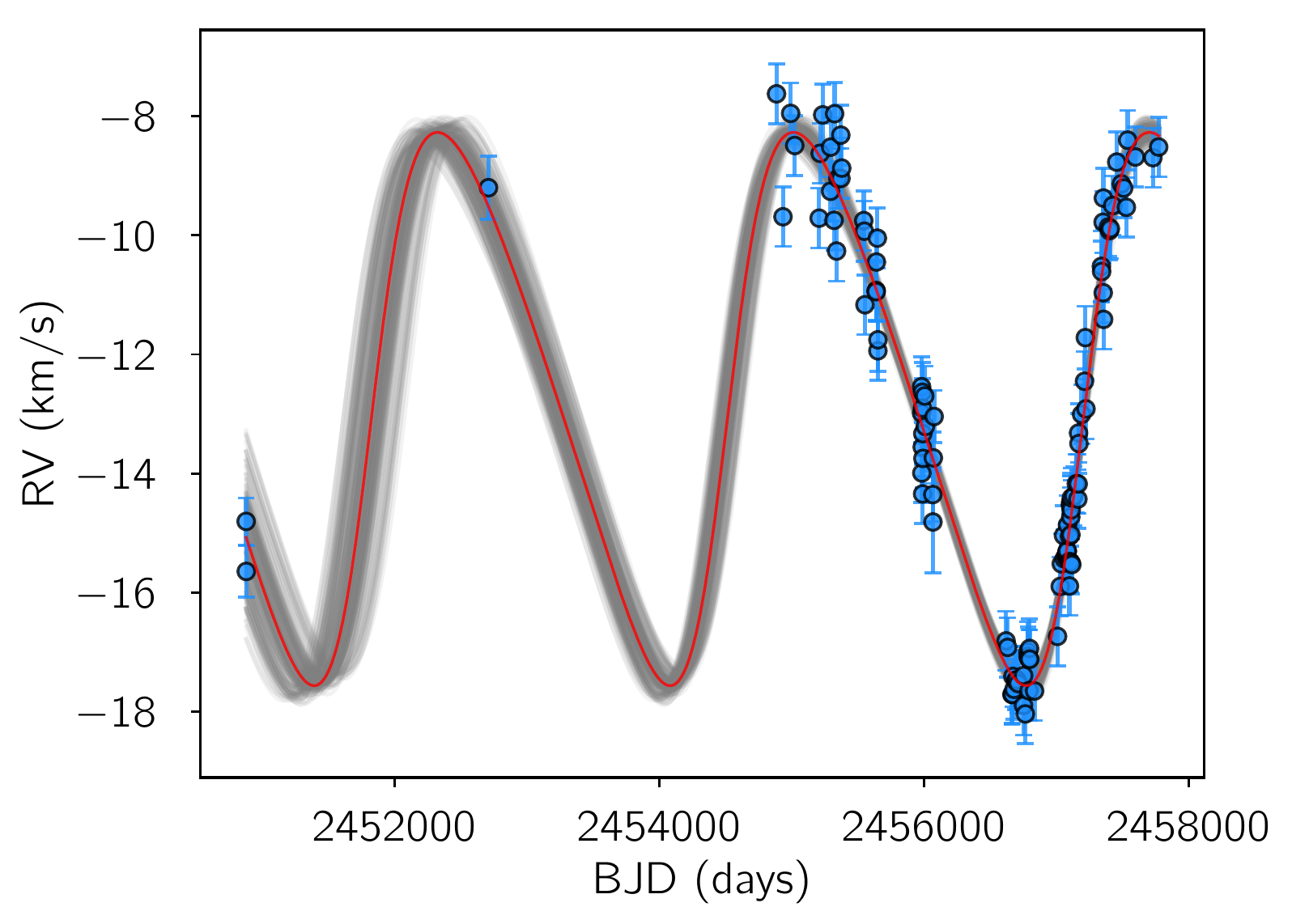}
\caption{Radial velocity of LoTr 5 in a 20 years time span covering two orbits of the long-period companion. The red line represents the median model from the posterior distribution of the fitted parameters (see text for details) and the gray lines represent a random subsample of models within the 68.7\% confidence interval.}
\label{fig:rv_long}
\end{figure}

In Figure~\ref{fig:rv_long}, we show the radial velocity data described in Sect.~\ref{sec:cafe}, covering 20 years time span and including ELODIE, HERMES and CAFE observations. They clearly show the long-term variations found in \cite{Jones2017}, corresponding to a $\sim$7.4-years period eccentric companion to the fast rotating G-type star.  For the first time, we report the coverage of two cycles in the orbit, with values in perfect agreement with the scenario proposed in \cite{Jones2017}. Given the new data, the Lomb-Scargle periodogram of the radial velocities shows a clear peak at $\sim$ 2700 days. We have modelled these radial velocity data by using six free parameters to account for the orbital and physical properties of the long-period companion: period ($P$), time of periastron passage ($T_{0}$), systemic velocity of the system (V$_{sys}$), radial velocity semi-amplitude ($K$), eccentricity ($e$) and argument of the periastron ($\omega$). Added to this, we included a radial velocity offset for each instruments pair ($N_{\rm inst}$-1 additional parameters) and a jitter term for each instrument ($N_{\rm inst}$ additional parameters) to account for possible white noise mainly due to underestimated uncertainties or unaccounted instrumental effects. So, in total, we fitted for 11 parameters.

We used the implementation of Goodman \& Weare's affine invariant Markov chain Monte Carlo (MCMC) ensemble sampler \textit{emcee}\footnote{See \url{http://dan.iel.fm/emcee/current/} for further documentation.}, developed by \cite{Foreman-mackey2013} to sample the posterior probability distribution of each of those parameters. We set uniform priors to all parameters in the ranges stated in Table~\ref{Table:priors_RV} and used 50 walkers with 5000 steps each. The posteriors are then computed by combining the latter half of all chains to avoid any possible dependency with the initial parameters. The chains converged quickly and show no clear correlations between the parameters.  In Table~\ref{Table:best_fit_RV} we show the median and 68.7\% confidence intervals for the different parameters. 

\begin{table*}
 \centering
  \caption{Best fit from the radial velocity analysis.}
  \begin{tabular}{lcccc}
  \hline    
Parameter	& 	Long-period  &  5.95d & $\sim$129d (circ) & $\sim$129d (ecc) \\
  \hline \hline
$V_{\rm sys}$ 	(km\,s$^{-1}$)	& 	-8.53$\pm$ 0.37  & -	& -0.009$\pm$0.090 &-0.020$\pm$0.080\\
$P$ 	(days)	&  	$2689 \pm 52$ & $5.96675\pm0.00051$ & $128.9^{+6.1}_{-5.7}$&129.8$\pm$1.6\\
$T_0$ 	(days)	&  	2455944$ \pm 25$ & - & -	& $2454976^{+50}_{-8}$\\
$K$ 	(km\,s$^{-1}$)	& 	4.630$\pm$ 0.084 & 0.609$\pm$0.028 & $0.089^{+0.107}_{-0.079}$& $0.279^{+0.028}_{-0.133}$	\\
$e$ 		&  	$0.249\pm0.018$ &- & 0.0	&$0.60^{+0.17}_{-0.31}$\\
$w$ ($^{\circ}$) 	&  259.9$\pm$4.8  & -   & - & $-26^{+55}_{-31}$\\
\hline
\end{tabular}
\label{Table:best_fit_RV}
\end{table*}

The median solution provides a period of  $2689 \pm 52$ days and an eccentricity of $0.249\pm0.018$, in agreement with those derived by \cite{Jones2017}. These values and their corresponding confidence intervals are listed in Table\,\ref{Table:best_fit_RV}. Assuming a mass for the (sub)giant star of 1.1 M$\odot$, we obtain a minimum mass for the companion of $m_2\sin{i} = 0.499~M_{\odot}$.  If we assume that the orbital plane of the binary is co-planar with the waist of the nebula, then the inclination angle of the system is $i=17^{\circ}$ \citep[see ][]{Graham2004} and the absolute mass for the companion would be $1.5 M_{\odot}$. This mass is incompatible with a white dwarf-like star. However, this is not very conclusive, since the orbital plane might not be coplanar with the waist of the nebula and/or the inclination of the system might be different than that derived from the model by \cite{Graham2004}. For example, assuming that the inclination angle of the system is i$\sim$45$^{\circ}$ (as proposed by \cite{Strassmeier1997}), a mass for the companion of $\sim$0.7\,M$\odot$ is obtained, which is clearly in the range of the central stars of PNe. Also, with smaller inclinations (>23$^{\circ}$) we obtain a white dwarf mass below the Chandrasekhar limit (1.4 M$\odot$).

However, if we assume coplanarity of the long-period orbit and the waist of the nebula, the mass of this second object would be outside of the typical masses for a PN progenitor. Consequently, as stated by \cite{Jones2017}, either this long-period companion is not coplanar with the waist of the nebula, or the inclination of the waist is much larger than the reported $17^{\circ}$, or the PN progenitor is blended in a close binary system forming a hierarchical triple. Thus, in the following section we further explore the radial velocity residuals to look for additional bodies/periodicities in the system.

%----------------------------------------------------
\subsubsection{The $\sim$5.9~days period}
%----------------------------------------------------

\begin{figure}
\includegraphics[width=\columnwidth]{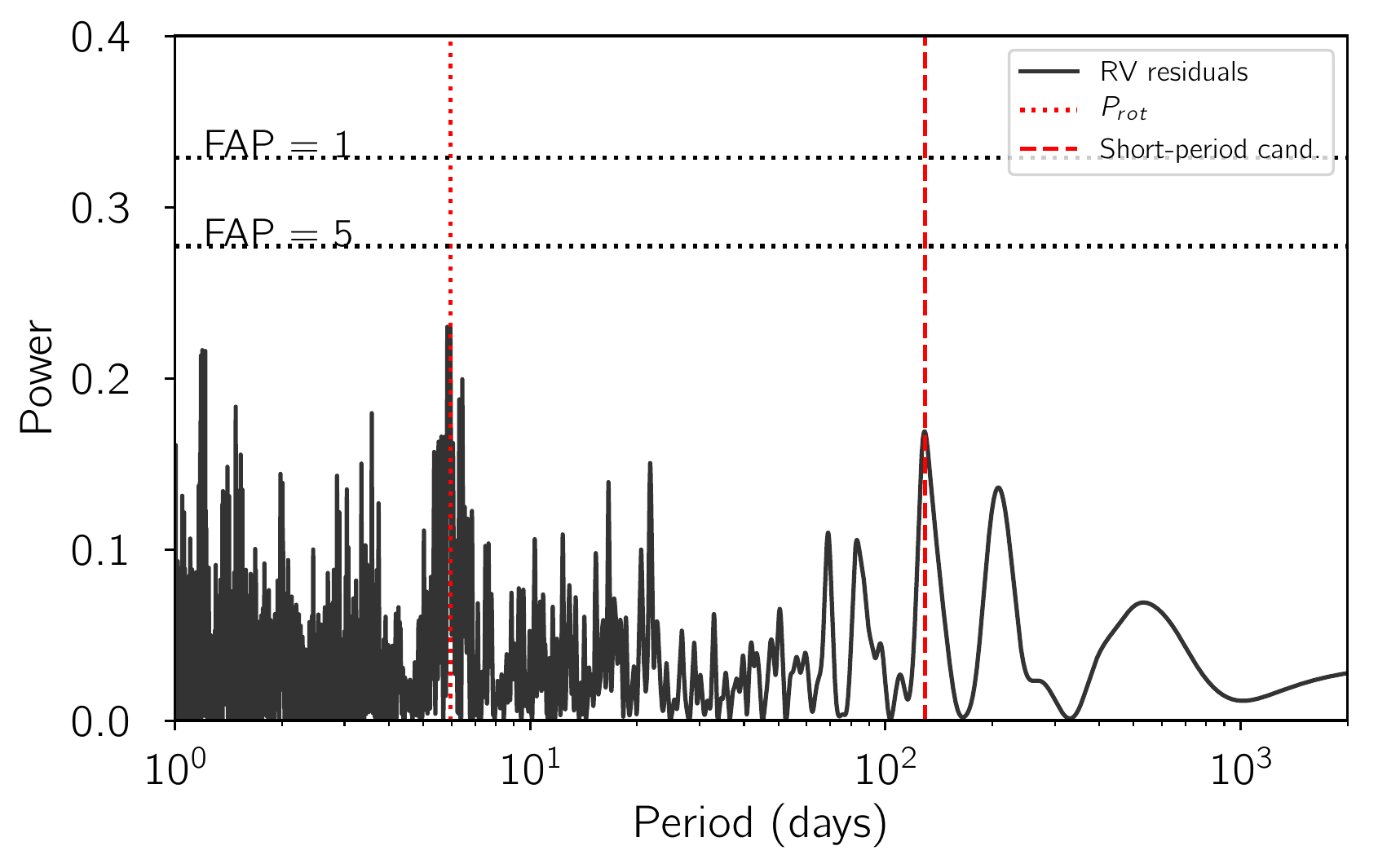}
\includegraphics[width=\columnwidth]{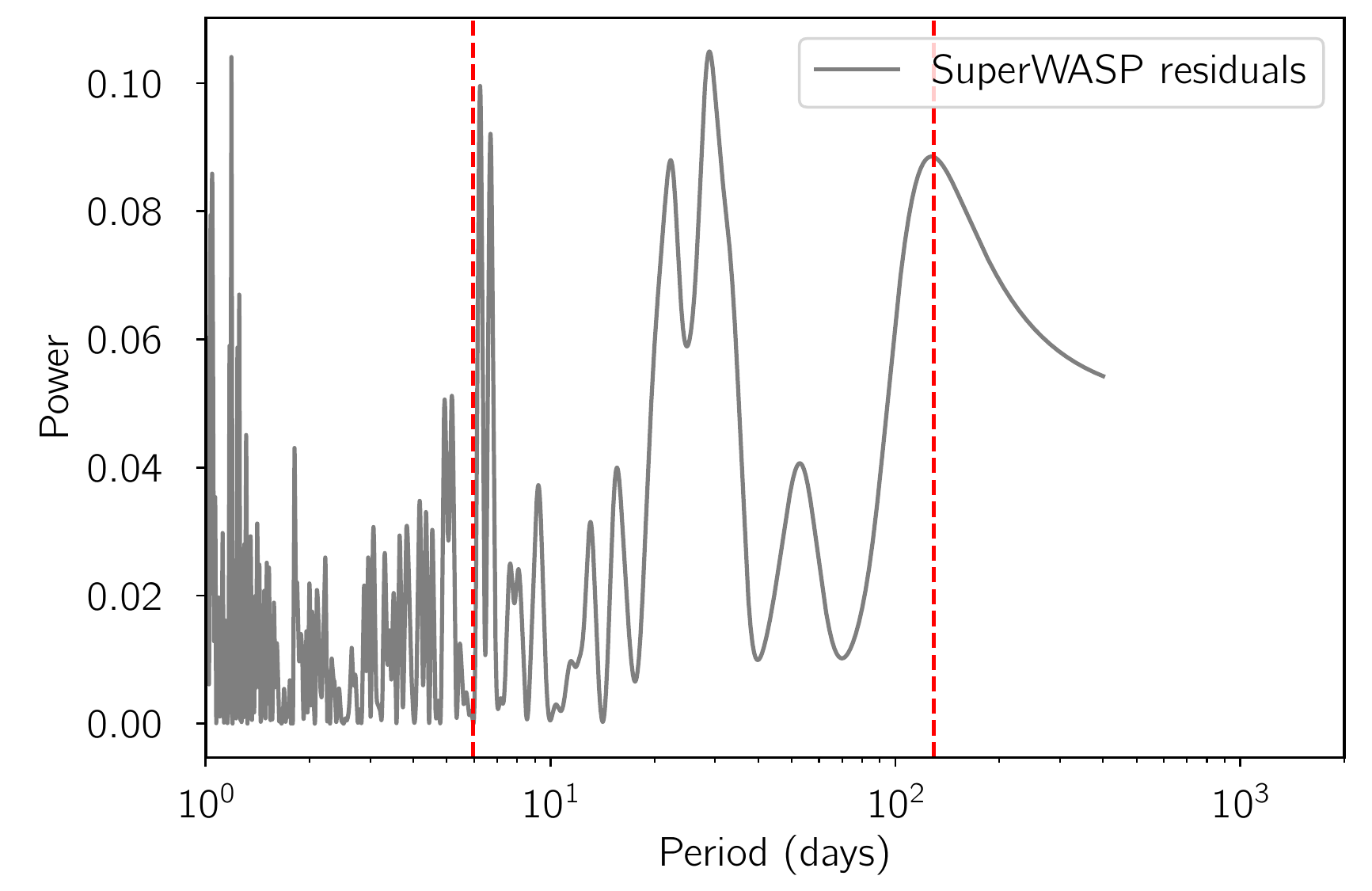}
\caption{(Top:) Lomb-Scargle periodogram of the residuals of the radial velocity data after removing the long-period component. The false alarm probabilities (FAP) at 1\% and 5\% are marked with dotted lines. (Bottom:) Lomb-Scargle periodogram of the SuperWASP dataset after removing the 5.9 days period component. The peak at 129 days is displayed in red dashed line.}
\label{fig:RVresidualsperiodogram}
\end{figure}

In Figure.~\ref{fig:RVresidualsperiodogram}, we show the Lomb-Scargle periodogram of the residuals of the radial velocity data from the long-term companion analysis. For this analysis, we have used the data from \cite{Jones2017} with JD $\ge$ 2456700, since they have a lower dispersion in comparison with the previous data. The strongest peak corresponds to $P\sim5.95$~days, previously identified in our light curve analysis. This periodicity could be either explained as stellar rotation-induced activity signals (as previously assumed) or due to a third guest in the system, i.e., a low-mass short-period companion to the fast rotating G-type star. We first investigate this latter scenario (although we note that this scenario would make worse the problem with the mass of the outer white dwarf) by fitting a circular Keplerian orbit to the radial velocity after removing the long-period signal. The results of this analysis are provided in the third column of Table~\ref{Table:best_fit_RV}. As we show, the period perfectly matches the significant periodicity found in the light curves. The amplitude of this assumed Keplerian orbit would be of the order of 600 m/s. Assuming the (sub)giant star has 1.1 $M_{\odot}$, the mass of the corresponding companion would be $m_3\sin{i}=5.8~M_{\rm Jup}$, i.e., in the planetary or brown dwarf domain for most inclinations. However, the light curve modulations (both the large amplitude, the presence of only one minimum per period and no eclipses), allows us to discard this configuration.

A second (much more likely) scenario is that this short-period radial velocity modulation is a consequence of the high chromospheric activity of the G-type star. This interpretation is supported by the photometric observations, which present sinusoidal variability with approximately the same period - this photometric variability is generally accepted to due to the presence of short-lived spots on the stellar surface which move with the giant's rotation period \citep{Miszalski2013b}.  Furthermore, the amplitude of these radial velocity variations, at roughly 1\% of the rotation velocity, is also consistent with amplitude of the photometric observations, at 1--2\% depending on the filter.  As such, this is the first time that the chromospheric activity of this star has been shown spectroscopically, principally due to the extremely high precision and good time coverage of the HERMES data.

%----------------------------------------------------
\subsubsection{The $\sim$129 days period}
\label{sec:129days}
%----------------------------------------------------

By inspecting the periodogram of the residuals of the radial velocity after removing the long-period signal (Figure.~\ref{fig:RVresidualsperiodogram}), the second highest peak is found at $\sim 129$~days. Despite not being statistically significant (false alarm probability larger than 5\%), we remark this peak because we also found relevant modulation with the exact same periodicity in the photometric data (see Sect.~\ref{sec:lc} above).

We tried to fit the residuals with another Keplerian model, now assuming circular orbit for the sake of simplicity of the adopted model. Assuming that the 5.9~days variation in radial velocity is due to the stellar activity observed in the light curves, we can include in our model a Gaussian Process (GP) with a quasi-periodic kernel (see \citealt{Faria2016}) to account for the correlated noise introduced by the stellar activity mentioned before. The hyper-parameters of this model include an amplitude of the variations ($\eta_1$), a correlation decay timescale ($\eta_2$), and a periodic component with period $\eta_3$. The last hyper-parameter ($\eta_4$) controls the weight of the two periodic components. The Keplerian model includes a residual systemic velocity of the system ($V_{\rm sys,3}$), the semi-amplitude of the radial velocity signal ($K_3$), the period of the additional component ($P_3$), and its time of conjunction ($T_{0,3}$). Added to these eight parameters, we also include a jitter parameter to account for possible white noise ($\sigma_{\rm jit}$). We provide uninformative priors for all of the parameters but restricting their ranges to particular regimes. In Table~\ref{Table:priors_RV}, we provide the priors.

The parameter space is explored by using the same approach explained in Sect.~\ref{sec:long}. Despite the uninformative priors, the MCMC chains converge into a solution corresponding to the periodicity found in the periodogram and in the superWASP light curve analysis. The median values for each parameter and their 68.7\% confidence interval are provided in Table~\ref{Table:best_fit_RV}. The radial velocity semi-amplitude parameter is, however, not constrained, and we can only provide an upper limit of $K_3<390$~m/s. Again assuming a 1.1$M_{\odot}$ for the primary star, this would imply a maximum projected mass of $m_3\sin{i}<10.3~M_{\rm Jup}$, providing an absolute mass in the case of coplanarity with the PN waist of $m_3 \sim 35.3~M_{\rm Jup}$. However, it is again really difficult to explain this peak in the light curve with such a low mass.

We have also explored the possibility of an eccentric orbit. In this case, we also obtain a median model in agreement with the $\sim$~129-days period. Interestingly, in this case, the posterior distribution of the eccentricity is very broad but favouring very high eccentricities ($e>0.5$), with the most probable value being around $e\sim 0.75$. We note that given the small amplitude of the signal and the rapid rotation of the star, the detection is somehow marginal but provides hints for the presence of a third guest in a highly eccentric orbit. This high eccentricity could be explained by the presence of the exterior massive companion (also in an eccentric orbit). Additionally, assuming $e=0.75$, this third guest would be as close to the primary G-type star as $r=0.12$~AU during the periastron passage. This could thus be the source of the variability that we detected in the superWASP light curve with the same periodicity. However, with the current data we are not able to discuss the origin of the light curve modulations (i.e., irradiation or tidal interactions).

\subsection{The puzzling H$\alpha$ double-peaked profile}

\begin{figure}
	\includegraphics[width=\columnwidth]{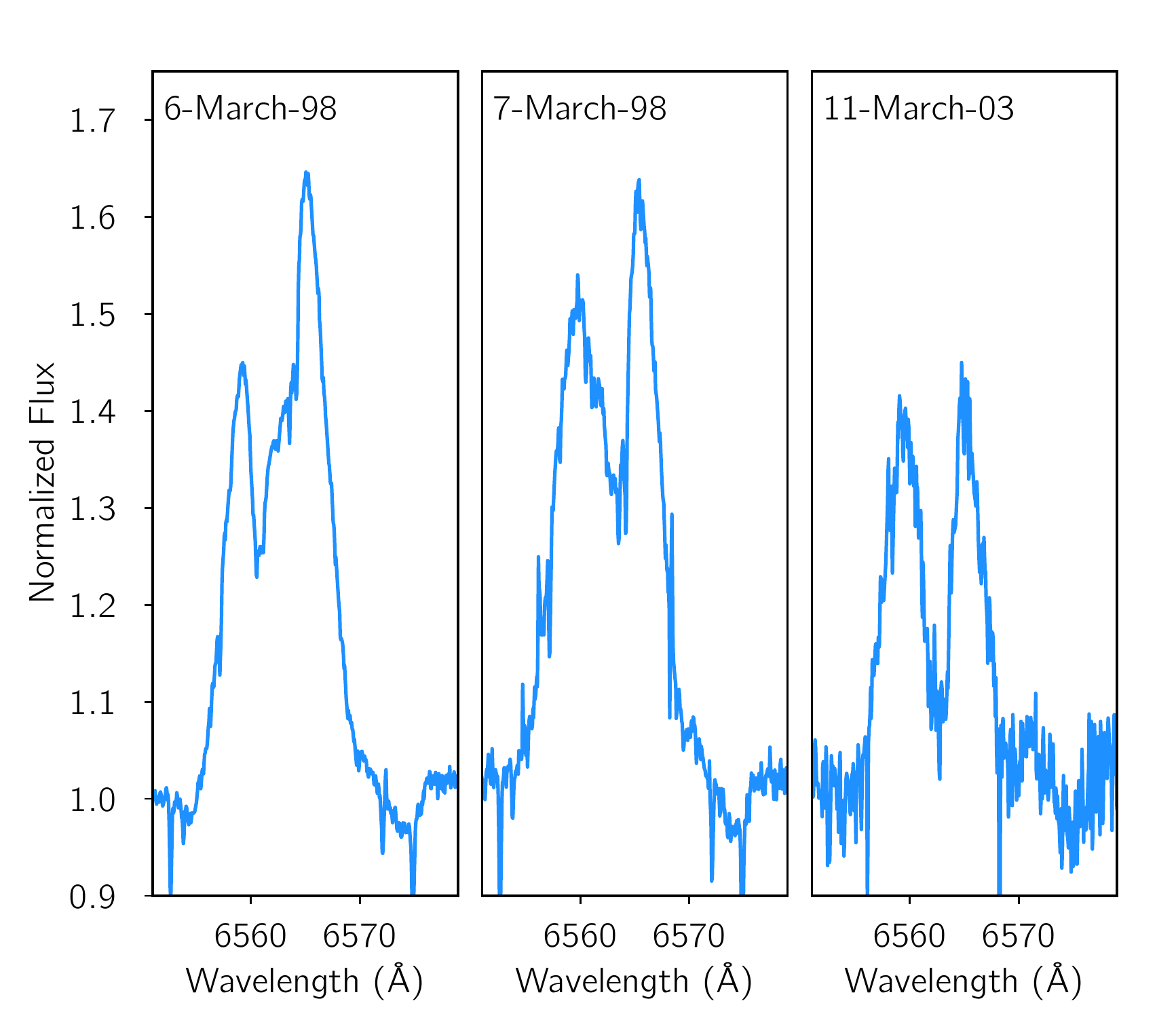}
	\includegraphics[width=\columnwidth]{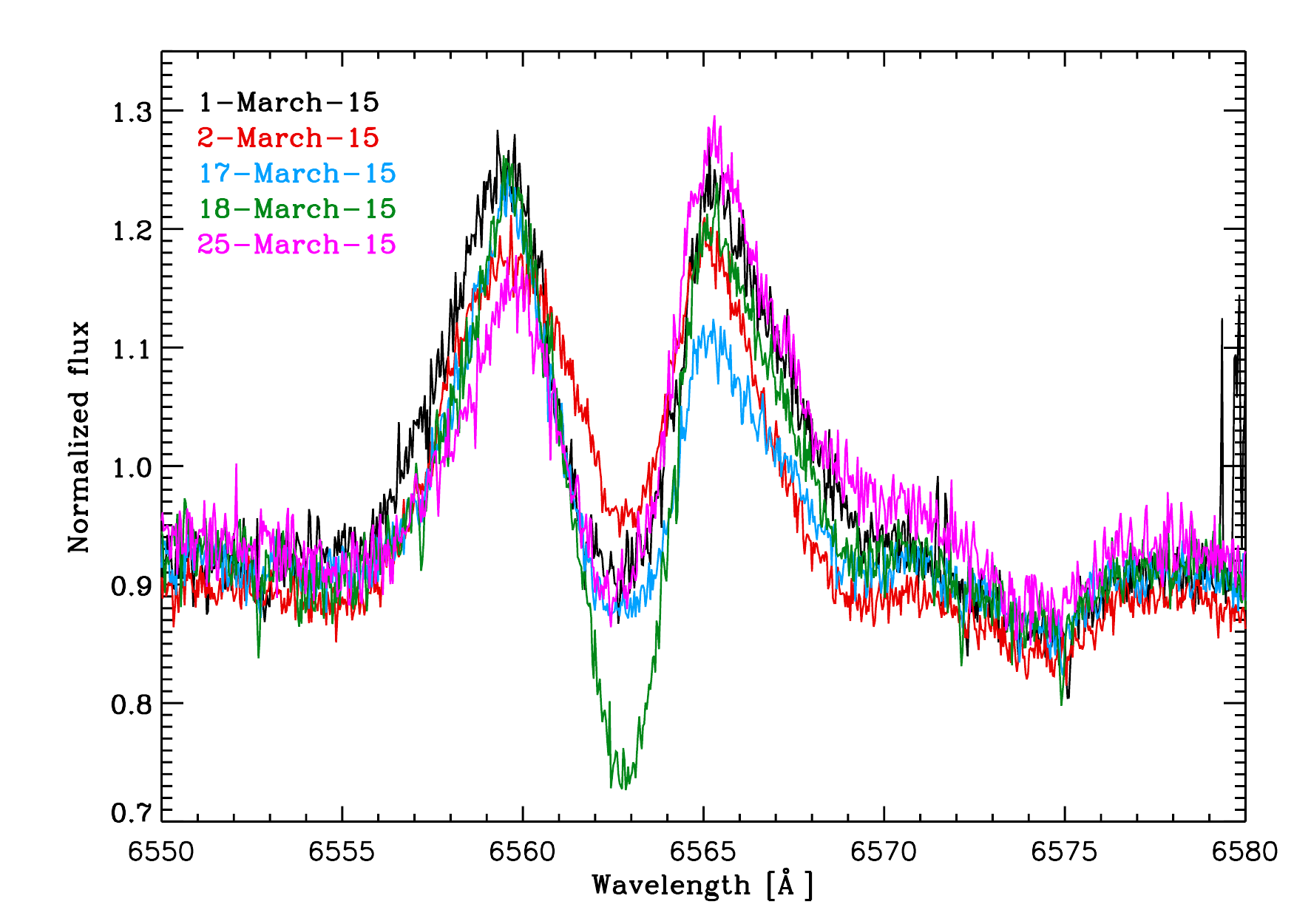}
	\includegraphics[width=0.991\columnwidth]{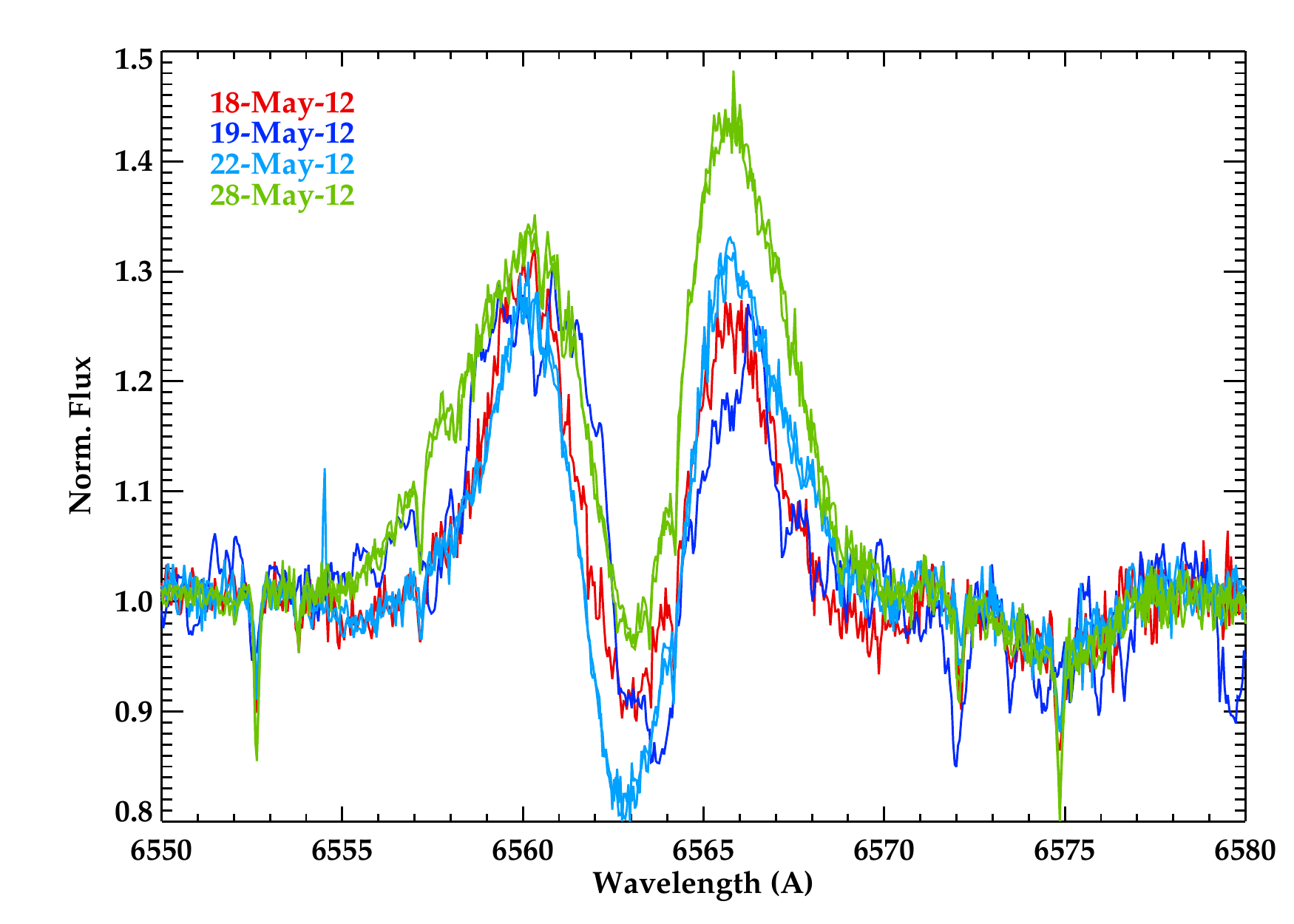}
    \caption{Variations in the H$\alpha$ profile during the ELODIE (top panel), HERMES (middle panel) and CAFE observations (bottom panel).}
    \label{fig:halpha}
\end{figure}

Figure\,\ref{fig:halpha} shows the rapid variations of the H$\alpha$ profile during the ELODIE (top panel), HERMES (middle, in a small sample) and CAFE (bottom) observations. The three datasets show clear H$\alpha$ double-peaked profile that varies with very short time scales, with a strong absorption feature which reaches below the continuum level in some cases. Such H$\alpha$ double-peaked profile is not new in LoTr 5 and was already reported by \cite{Jasniewicz1994} and \cite{Strassmeier1997}. It has also been seen in other systems like, e.g., LoTr 1 \citep{Tyndall2013} and Abell\,35 \citep{Acker-Jasniewicz1990, Jasniewicz1992} although the origin is still unknown. Several explanations have been proposed, specially the chromospheric activity of the G-type star, the presence of an accretion disc or the existence of strong stellar winds. In the following, we will discuss these three possibilities in more detail.

%----------------------------------------------------
\subsubsection{Chromospheric activity}
%----------------------------------------------------

\begin{figure}
	\includegraphics[width=\columnwidth]{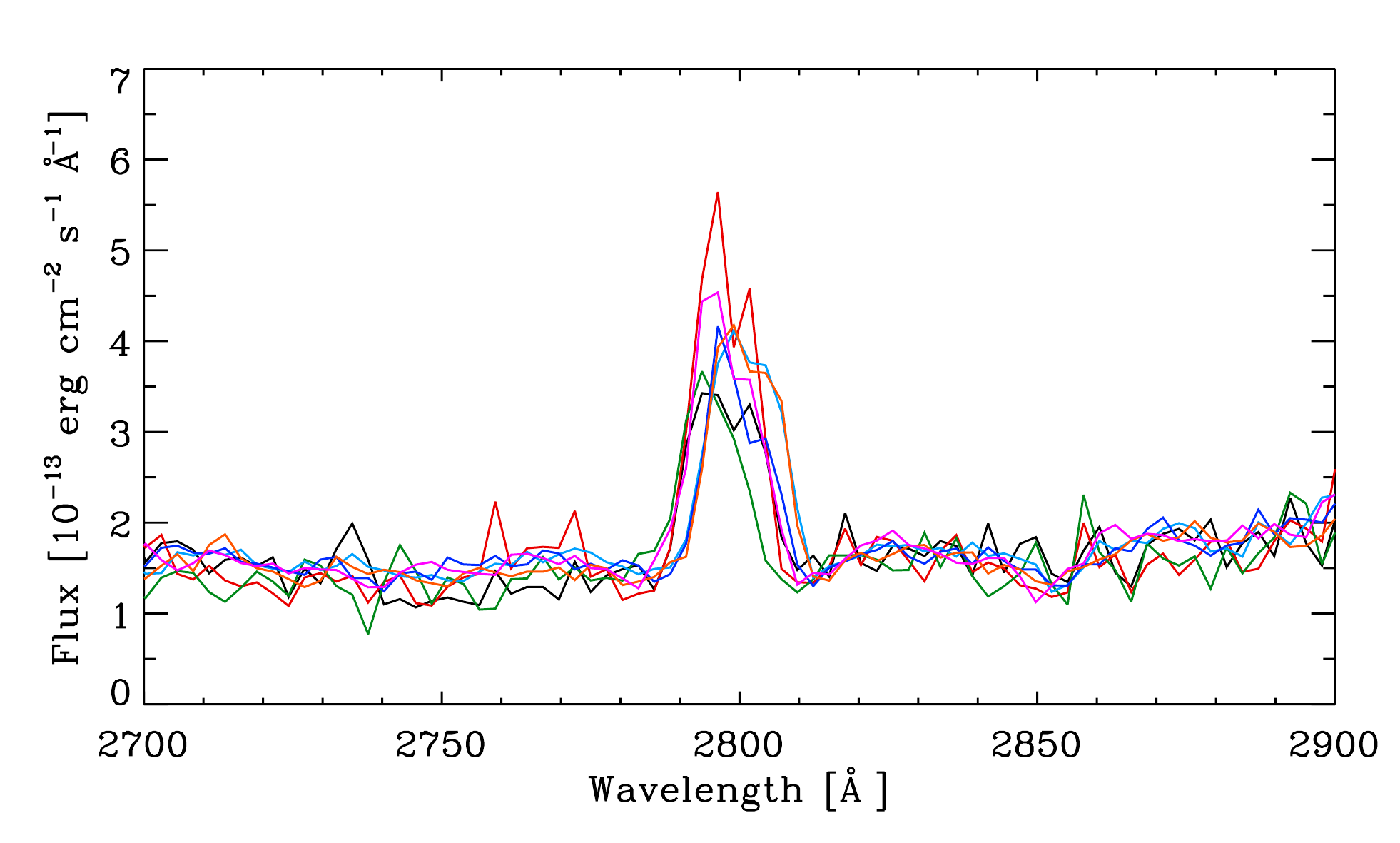}
    \caption{Variations in the MgII\,$\lambda$2800 emission line in several IUE spectra.}
    \label{fig:MgII_line}
\end{figure}

There is no doubt that the G-type star of HD\,112313 has a high chromospheric activity. This is clearly indicated by the active-chromosphere indicators present in the spectrum. \cite{Strassmeier1997} already showed some of these relevant indicators like the emission in the Ca\,{\sc ii} H\&K lines, the infrared triplet lines of Ca\,{\sc ii}\,$\lambda$8498,8542,8662 and the already mentioned broad, double-peaked H$\alpha$ profile. Also, the Mg\,{\sc ii} h\&k \,$\lambda$2796,2803 lines appear in emission in the IUE spectra, as firstly shown by \cite{Feibelman-Kaler1983}. For the first time, we find fast variations of this line in the IUE spectra (see Figure\,\ref{fig:MgII_line}), in contrast  to what \cite{Jasniewicz1996} reported. Finally, it should not be forgotten the detection of X-rays emission in LoTr 5 \citep{Apparao1992,Montez2010}. According to \cite{Montez2010} the most likely explanation for this X-rays detection is the presence of coronal activity associated to the G star. All of these pieces seem to indicate that chromospheric activity plays an important role in LoTr 5. However, some features of the H$\alpha$ profile remain unexplained. The most relevant is the full width at half maximum (FWHM) of the line, that is too high (450 km\,s$^{-1}$) to be due only to chromospheric activity. Also, if we measure the equivalent width (EW) of the H$\alpha$ core emission (determined by subtracting an inactive star with the same effective temperature and surface gravity) it appears to be quite higher than that expected in the (sub)giant regime, according to Figure\,5c in \cite{Strassmeier1990}. In contrast, the EW is similar to that measured in FK Com, another rapidly rotating giant star that is known to have high chromospheric activity. FK Com presents an H$\alpha$ double-peaked profile  similar than HD\,112313 \citep[see, for instance,][]{Kjurkchieva-Marchev2005}. However, we note that, in contrast to what occurs in FK Com, in LoTr 5 (i) there is no evidence of an accretion disc via RLOF (see \ref{subsubsection:accretion_disk})  and (ii) there is not any phase dependency in the variation of the H$\alpha$ profile (see below). All this would point out that part of the H$\alpha$ line is not due to chromospheric activity.

Following the procedure explained in \cite{Acker-Jasniewicz1990} for Abell\,35, we have tried to find a modulation of the H$\alpha$ profile with the rotation phase, by measuring the flux variation between the nearby continuum and the deepest point of the H$\alpha$ absorption core. In contrast to the results found by \cite{Acker-Jasniewicz1990} in the case of Abell\,35, we do not find any correlation with the rotational period during the long time span of the observations in LoTr 5. We have inspected the H$\alpha$ profile in all the Hermes spectra and we found no correlation with the rotation period, see Fig.\,\ref{fig:trailed}, where we show the phased-folded trailed spectra of the H$\alpha$ line with the rotation period.

The Ca\,{\sc ii} H\&K $\lambda$3933,3967 lines appear in emission in the HERMES dataset (as already reported by \citealt{Strassmeier1997}). We do not find variations in these lines at the level of the H$\alpha$ variations that we find in HERMES data. Hence, Ca\,{\sc ii} H\&K lines are not useful for studying the activity of the star.

%----------------------------------------------------
\subsubsection{Accretion disc around the possible close companion}
\label{subsubsection:accretion_disk}
%----------------------------------------------------

The presence of an accretion disc in LoTr 5 has been also discussed in the literature. If indeed the H$\alpha$ emission was due to an accretion disk, then the double-peaked profile would come from two different emission regions instead of a broad emission feature. \cite{Montez2010} already discussed the system within this scenario: since the presence of an accretion disc would imply the Roche lobe overflow (RLOF), they calculated the Roche lobe radius (R$_{L}$) assuming the synchronization of the rotation and orbital period (i.e., they assumed a close 0.6M$\odot$ companion at $\sim$5.9 days). Under this configuration, the R$_{L}$ was below the radius of the star, filling its Roche Lobe and allowing the existence of an accretion disc. However, from our radial velocity analysis, we know that it is not possible to have such a massive companion so close of the G-type star. On the contrary, in the case we would have a low-mass object at $\sim$5.9 days, the R$_{L}$ would be $\sim$16R$\odot$, and the star could not have filled its Roche lobe. Therefore, with the current data, we conclude that it is not possible to have an accretion disc via RLOF in LoTr 5. 

Finally, and as mentioned in Section\,\ref{section:SED}, no significant infrared excess is identified in the SED, which would point out that an accretion disc is highly unlikely in the system.

%----------------------------------------------------
\subsubsection{The possible resemblance with symbiotic stars}
%----------------------------------------------------

Finally, it is worth mentioning that this double-peaked H$\alpha$ emission resembles the observed in some symbiotic stars (SS), the widest interacting binary systems known with a white dwarf accreting material from a giant star via stellar winds \citep[see, e.g.,][]{VanWinckel1993,Burmeister-Leedjarv2009}. These systems are sometimes embedded into a bipolar nebula similar to some PNe (like LoTr 5). However, apart from the complex H$\alpha$ profile, we do not find evidence of P-cygni profiles and/or strong complex emissions in any of the lines in the spectra of HD\,112313. The only detection of P-cygni profiles in HD\,112313 was reported by \cite{Modigliani1993}, who found those profiles in the O\,{\sc v}\,$\lambda$1371 and C\,{\sc iv}\,$\lambda$1548,1550 lines in the IUE spectra. From the analysis of these profiles, they concluded a fast wind speed of 3300 km\,s$^{-1}$ (although we note that, if true, this fast wind more likely would come from the hot central star and not from the cool giant). In any case, we have re-analyzed the IUE spectrum used by \cite{Modigliani1993} for HD\,112313 (SWP19909), as well as the rest of them -both in low- and high-resolution- and have not found any evidence of P-cygni profiles. All these spectra were extracted from the INES database; in order to double check the results, we have also retrieved the IUE spectra available from MAST\footnote{\url{https://archive.stsci.edu/iue/search.php}}, that are reduced in an independent way to that of the INES archive. No P-cygni profiles are identified in MAST spectra either. The absence of these characteristic profiles and other prominent features typical in symbiotic stars make it highly improbable that LoTr 5 belongs to this group.

With a high-quality spectrum of the nebula we would be able to place LoTr 5 in a diagnostic diagram \citep[like those shown in][]{Ilkiewicz-Mikolajewska2017} to distinguish between symbiotic stars and planetary nebulae. Unfortunately, only a very weak H$\alpha$ flux \citep[by][]{Frew2016} and [O\,{\sc iii}]\,$\lambda$4959,5007  \citep[measured by][]{Kaler1990} are published in the literature.

\section{Conclusions and discussion}

We revisit the complex LoTr 5 system and discuss the still open questions around this object by means of radial velocity and archival (SuperWASP, ASAS and OMC) light curve analysis. For the first time, and based on both archive and new radial velocity data, we recover two orbital cycles of the long-period binary central star, confirming the orbital period of 2689 $\pm$ 52 days with an eccentricity of 0.249 $\pm$ 0.018. After removing this long-period component from the radial velocity, the residual periodogram shows two interesting (although still not statistically significant) peaks: the first one, at $\sim$5.9 days, is also identified in the SuperWASP and ASAS periodograms. This periodicity was already known and is assumed to be the rotation period of the G-type subgiant; the second peak, at $\sim$129 days,
albeit with low S/N, is also found when subtracting the 5.9 days component from the SuperWASP data (the most precise light curve we have). We analyze the $\sim$129 days periodicity and found that it could be compatible with a low-mass object (in the planetary or brown dwarf domain) with a high eccentricity. However, the current data are not enough to confirm such a third guest in the system and more dedicated data are mandatory to prove this scenario. 

The possibility of a triple system is also discussed under other configuration: assuming a third object with the same orbital period as the rotation period (5.9 days), i.e., assuming synchronization of the rotation period of the G-type star and the orbital period of a possible close binary. However, it appears unlikely with the current data since the most plausible explanation for the 5.9-days periodicity found in the radial velocity and photometric data is, indeed, the activity of the G-type star.

Finally, and regarding the chromospheric activity, we also discuss the double-peaked H$\alpha$ profile seen in the spectra, which shows moderate variations in very short time span. It is clear that the G-type stellar component of HD\,112313 has a high chromospheric activity. This is confirmed by the active-chromosphere indicators present in the UV and optical spectra, like the emission in the Ca\,{\sc ii} H\&K lines, the infrared triplet lines of Ca\,{\sc ii}\,$\lambda$8498,8542,8662, that appear less deep than in a non-active G-star with the same effective temperature and surface gravity (i.e., an emission component is clearly contributing to these absorption lines), as well as the Mg\,{\sc ii} h\&k \,$\lambda$2796,2803 lines, that appear in emission in the IUE spectra and show clear variations in intensity. However, some features of the complex H$\alpha$ profile appear not compatible with chromospheric activity (the most relevant is the high FWHM of the line), pointing out that at least part of the line is not due to chromospheric activity, and suggesting other possible explanations like the presence of an accretion disk and/or stellar winds. Based on the calculation of the Roche lobe radius, we discard the presence of an accretion disk via Roche lobe overflow, since the star can not have filled its Roche lobe. In addition, the absence of P-cygni profiles in the spectra of LoTr 5 points out that stellar winds do not play an important role in this system and, therefore, that we are not dealing with a symbiotic star.

\section*{Acknowledgements}

We thank our anonymous referee for his/her comments that have improved the discussion of the data. A.A. acknowledges support from FONDECYT through postdoctoral grant 3160364. LFM acknowledges partial support from Spanish MINECO grant AYA2014-57369-C3-3-P (co-funded by FEDER funds). Based on observations obtained with the HERMES spectrograph, which is supported by the Research Foundation - Flanders (FWO), Belgium, the Research Council of KU Leuven, Belgium, the Fonds National de la Recherche Scientifique (F.R.S.-FNRS), Belgium, the Royal Observatory of Belgium, the Observatoire de Genve, Switzerland and the ThŸringer Landessternwarte Tautenburg, Germany.  Based on observations collected at the German-Spanish Astronomical Center, Calar Alto, jointly operated by the Max-Planck-Institut fŸr Astronomie (Heidelberg) and
the Instituto de Astrof'sica de Andaluc'a (CSIC). Based on spectral data retrieved from the ELODIE archive at Observatoire de Haute-Provence (OHP). This publication makes use of VOSA, developed under the Spanish Virtual Observatory project supported from the Spanish MICINN through grant AyA2011-24052. Based on data from the OMC Archive at CAB (INTA-CSIC), pre-processed by ISDC. This paper makes use of data from the DR1 of the WASP data (Butters et al. 2010) as provided by the WASP consortium, and the computing and storage facilities at the CERIT Scientific Cloud, reg. no. CZ.1.05/3.2.00/08.0144 which is operated by Masaryk University, Czech Republic. This work used the python packages numpy, matplotlib, astroML, asciitable, pyTransit, kplr, and emcee.

%%%%%%%%%%%%%%%%%%%%%%%%%%%%%%%%%%%%%%%%%%%%%%%%%%

%%%%%%%%%%%%%%%%%%%% REFERENCES %%%%%%%%%%%%%%%%%%

% The best way to enter references is to use BibTeX:

\bibliographystyle{mnras}
\bibliography{Bibliography} % if your bibtex file is called example.bib

\appendix
\section{}

 \begin{table*}
 \centering
  \caption{Priors adopted for the analysis of the different radial velocity components and scenarios. $\mathcal{U}(a,b)$ stands for a uniform prior between a and b;  $\mathcal{G}(\mu,\sigma)$ are Gaussian priors with a mean $\mu$ and standard deviations $\sigma$; and $\mathcal{MJ}(a,b)$ is a Modified Jeffries prior with a and b parameters. }
  \begin{tabular}{lcccc}
  \hline    
Parameter	& 	Long-period  &  5.95d & $\sim$129d (circ) & $\sim$129d (ecc) \\
  \hline \hline
$V_{\rm sys}$ 	(km/s)	& 	$\mathcal{U}(-30,30)$  			& -		& $\mathcal{U}(-10,10)$ 		&$\mathcal{U}(-10,10)$ 	\\
$P$ 	(days)			&  	$\mathcal{U}(2000,4000)$ 		& X 		& $\mathcal{U}(115,140)$		&$\mathcal{U}(115,140)$	\\
$T_0$ 	(days)		&  	$\mathcal{U}(2455000,2459000)$ 	& - 		& -						& $\mathcal{U}(2454950,2455080)$\\
$K$ 	(km/s)			& 	$\mathcal{MJ}(0.005,15)$ 			& X 		& $\mathcal{MJ}(0.005,15)$	& $\mathcal{MJ}(0.005,3)$	\\
$e$ 					&  	$\mathcal{U}(0,1)$ 				&- 		& -						&$\mathcal{U}(0,1)$\\
$\omega$ ($^{\circ}$) 		&  $\mathcal{U}(0,360)$  				& -   		& - 						& $\mathcal{U}(0,360)$\\
$\eta_1$ (dex)			& -								&-		& $\mathcal{U}(-10.6,-6.9)$	& $\mathcal{U}(-10.6,-6.9)$ \\
$\eta_2$ (days)			& -								&-		&$\mathcal{U}(6,200)$		&$\mathcal{U}(6,200)$	 \\
$\eta_3$ (days)			& -								&-		&$\mathcal{G}(5.95,0.1)$		&$\mathcal{G}(5.95,0.1)$	 \\
$\eta_4$ 				& -								&-		&$\mathcal{U}(-10,20)$		&$\mathcal{U}(-10,20)$	 \\

\hline
\end{tabular}
\label{Table:priors_RV}
\end{table*}

\begin{figure*}
\includegraphics[width=\columnwidth]{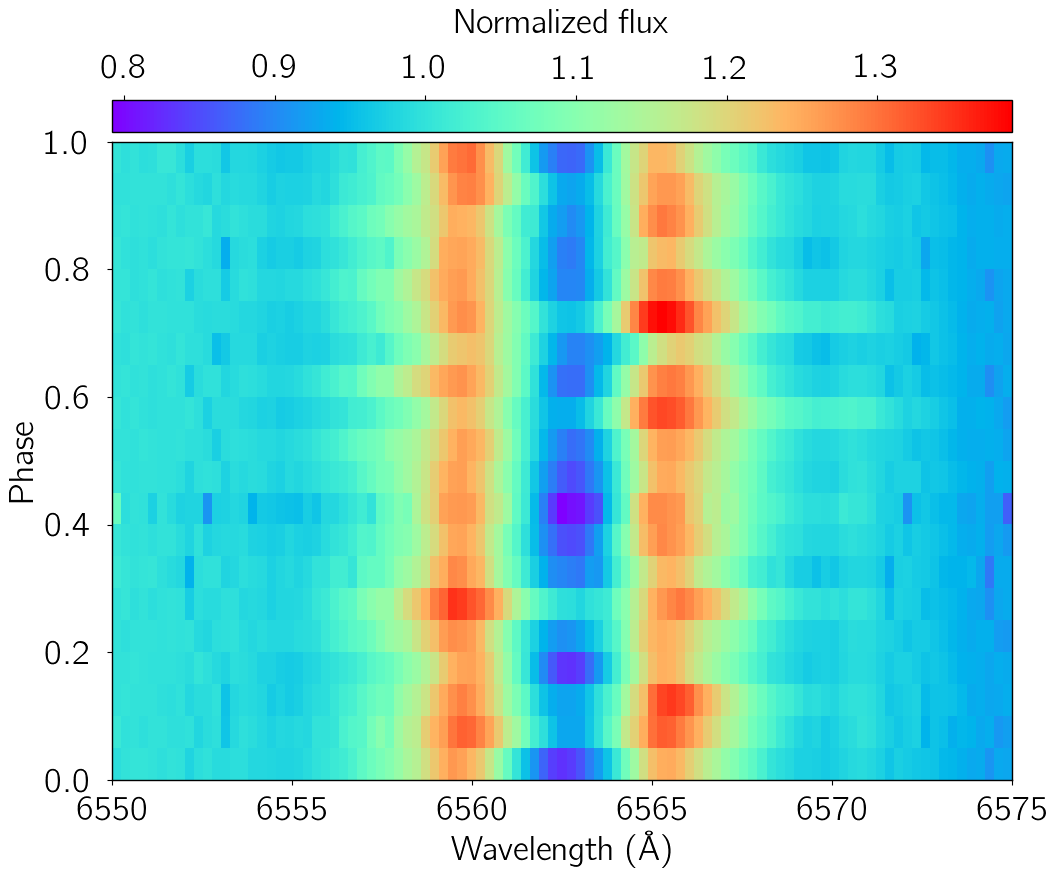}
\caption{Phase-folded trailed spectra of the H$\alpha$ line with the rotation period ($\sim$ 5.9 days) of the G-type star. No correlation is identified with such period, pointing out variations in the double-peaked H$\alpha$ profile are not only consequence of the chromospheric activity of the G-type star.}
\label{fig:trailed}
\end{figure*}

% Alternatively you could enter them by hand, like this:
% This method is tedious and prone to error if you have lots of references
%\begin{thebibliography}{99}
%\bibitem[\protect\citeauthoryear{Author}{2012}]{Author2012}
%Author A.~N., 2013, Journal of Improbable Astronomy, 1, 1
%\bibitem[\protect\citeauthoryear{Others}{2013}]{Others2013}
%Others S., 2012, Journal of Interesting Stuff, 17, 198
%\end{thebibliography}

%%%%%%%%%%%%%%%%%%%%%%%%%%%%%%%%%%%%%%%%%%%%%%%%%%

%%%%%%%%%%%%%%%%% APPENDICES %%%%%%%%%%%%%%%%%%%%%

%\appendix
%
%\section{Some extra material}
%
%If you want to present additional material which would interrupt the flow of the main paper,
%it can be placed in an Appendix which appears after the list of references.

%%%%%%%%%%%%%%%%%%%%%%%%%%%%%%%%%%%%%%%%%%%%%%%%%%

% Don't change these lines
\bsp	% typesetting comment
\label{lastpage}
\end{document}